\documentclass[fleqn,usenatbib]{mnras}  
\usepackage{newtxtext,newtxmath}
\usepackage{xfrac}
\usepackage{color}

\usepackage{ae,aecompl}
\usepackage[outline]{contour}
\usepackage{tikz}
\usepackage{algorithm}
\usepackage{bbold}

\newcommand{\fett}[1]{\boldsymbol{#1}}
\newcommand{\dd}{{\rm{d}}}

\newcommand{\be}{\begin{equation}}
\newcommand{\ee}{\end{equation}}
\newcommand{\RT}{\mathfrak{R}_\tau}
\newcommand{\xofq}{\text{\small $x(q,\tau)$}}
\newcommand{\xofqtext}{\text{\footnotesize $x(q,\tau)$}}
\newcommand{\xofqtextZA}{\text{\footnotesize $x_{\rm ZA}(q,\tau)$}}
\newcommand{\xofqtextZAstar}{\text{\footnotesize $x_{\rm ZA}(q_\star,\tau_\star)$}}

\newcommand{\xofqtextPSC}{\text{\footnotesize $x_{\rm PSC}(q,\tau)$}}

\newcommand{\xiPSC}{\xi_{\text{\fontsize{6}{7}\selectfont\rm PSC}}}
\newcommand{\PSC}{{\text{\fontsize{6}{7}\selectfont\rm PSC}}}
\newcommand{\PPSC}{{\text{\fontsize{6}{7}\selectfont\rm PPSC}}}
\newcommand{\PPPSC}{{\text{\fontsize{6}{7}\selectfont\rm PPPSC}}}
\newcommand{\ZA}{{\text{\fontsize{6}{7}\selectfont\rm ZA}}}
\newcommand{\xofqtextZAprime}{\text{\footnotesize $x_{\rm ZA}(q',\tau)$}}

\newcommand{\xofqprime}{\text{\small $x(q',\tau)$}}
\definecolor{darkgreen}{rgb}{0,0.5,0}

\definecolor{lime}{HTML}{A6CE39}
\DeclareRobustCommand{\orcidicon}{
	\begin{tikzpicture}
	\draw[lime, fill=lime] (0,0) 
	circle [radius=0.14] 
	node[white] {{\fontfamily{qag}\selectfont \tiny ID}};
	\draw[white, fill=white] (-0.0625,0.095) 
	circle [radius=0.007];
	\end{tikzpicture}
	\hspace{-2mm}
}

\foreach \x in {A, ..., Z}{%
	\expandafter\xdef\csname orcid\x\endcsname{\noexpand\href{https://orcid.org/\csname orcidauthor\x\endcsname}{\noexpand\orcidicon}}
}

\contourlength{0.2pt} 
\contournumber{2} 

\title[Singularities in the large-scale structure]{Unveiling the singular dynamics in the cosmic large-scale structure}

\author[Rampf, Frisch \& Hahn]{Cornelius Rampf$^{\,{\text{\tiny\orcidA{}}}\,\,\hyperlink{OCA}{1}}$\thanks{E-mail: \href{mailto:cornelius.rampf@oca.eu}{cornelius.rampf@oca.eu}}, Uriel Frisch$^{\,{\text{\tiny\orcidB{}}}\,\,\hyperlink{OCA}{1}}$ and Oliver Hahn$^{{\text{\tiny\orcidC{}}}\,\,\hyperlink{OCA}{1}, \hyperlink{AstroVienna}{2}, \hyperlink{MathVienna}{3}}$  \hypertarget{OCA} \\
$^{1}$Universit\'e C\^ote d'Azur, Observatoire de la C\^ote d'Azur, CNRS, Laboratoire Lagrange, Boulevard de l'Observatoire, CS 34229, 06304 Nice, France  \hypertarget{AstroVienna} \\
$^{2}$Department of Astrophysics, University of Vienna, T\"urkenschanzstraße 17, 1180 Vienna, Austria  \hypertarget{MathVienna}\\
$^{3}$Department of Mathematics, University of Vienna, Oskar-Morgenstern-Platz 1, 1090 Vienna, Austria}

\date{Accepted 19 May 2021. Received 24 April 2021; in original form 29 January 2021}

\pubyear{2021}

\begin{document}
\label{firstpage}
\pagerange{\hyperlink{firstpage}{L}\pageref{firstpage}--\hyperlink{lastpage}{L}\pageref{lastpage}}
\maketitle

\begin{abstract}
It is known that the gravitational collapse of cold dark matter leads to infinite-density caustics that seed the primordial dark-matter halos in the large-scale structure. The development of these caustics begins, generically, as an almost one-dimensional phenomenon with the formation of pancakes. Focusing on the one-dimensional case, we identify a landscape of non-differentiable, and thus, singular features in the particle acceleration that emerge after the first crossing of particle trajectories. We complement our fully analytical studies by high-resolution simulations and find outstanding agreement, particularly shortly after the first crossing. We develop the methods in one space dimension but outline briefly the necessary steps for the 3D case.
\end{abstract}


\section{Introduction}

A simple example of a singularity is when a function $\xi(\tau)$ 
has local behaviour $\propto (\tau - \tau_1)^\lambda$ around $\tau_1$, 
where $\lambda  \notin \mathbb{N}$ is the singularity exponent. 
If $\lambda$ is a negative integer, then the singularity has a simple 
pole-like structure. If $\lambda$ is instead a 
{\it positive non-integer number}, then certain derivatives of $\xi(\tau)$ 
will blow up around~$\tau_1$ and, as a result, $\xi(\tau)$ cannot be 
represented locally by a Taylor series.

In cosmology, the obvious singularities are density caustics that 
comprise the central building block for the cosmic large-scale structure.  
At the particle level, infinite-density caustics result from shell-crossing, 
the crossing of cold dark matter (CDM) trajectories.  Once particles have 
crossed for the first time, the single-stream flow becomes multi-stream. 
Subsequently, secondary gravitational infall commences, inducing more 
shell-crossings that lead to a proliferation of streams, and eventually 
to virialized structures.

Some of the singularities were classified by~\cite{Arnold1980,ASZ:1982}, 
by  exploiting an approximate nonlinear theory of gravitational instability, 
the Zel'dovich approximation (ZA; \citealt{Zeldovich:1969sb}).  However, 
singularities in derivatives of the particle trajectories, 
which we report here,  remained undetected as the ZA is an 
\textit{acceleration-free} model for nonlinear collapse, 
thereby being effectively blind to secondary gravitational infall.

Central to the analysis of~\cite{Arnold1980,ASZ:1982}, as well as
ours, is the use of Lagrangian-coordinates approaches to gravitational
instability that permit investigating singularities in a tractable
manner.  The ZA is the lowest-order Lagrangian-coordinates
approximation to the cosmological fluid equations (the single-stream
case of the Vlasov--Poisson equations). Furthermore, it is exact
in 1D~\citep{Novikov:2010ta}, as long as multi-stream flow has not yet
appeared.  Beyond 1D, higher-order approximations should be
incorporated, and the corresponding framework is dubbed Lagrangian
perturbation theory~\citep[LPT;][]
 {Buchert:1987xy,Bouchet:1992uh,Buchert:1992ya,
  EhlersBuchert97,Bernardeau:2001qr,Rampf:2012xa,Zheligovsky:2013eca}.
In LPT the displacement field is
the only dynamical variable, 
which is expanded as a (temporal or spatial) Taylor series.
Recently, the first nontrivial shell-crossing solutions in LPT have been 
identified~\citep{Rampf:2017jan,Rampf:2017tne}, while numerical evidence 
of convergence of LPT in 3D at shell-crossing was given 
by~\cite{Saga:2018nud,2021MNRAS.501L..71R}.

Nonetheless, the standard implementation of LPT cannot predict secondary 
gravitational infall, since it is based on
a fluid description that does not incorporate multi-streaming. 
Instead, the multi-streaming evolution of CDM is 
governed by the Vlasov--Poisson equations.
Following in the footsteps of~\cite{Colombi:2014lda,Taruya:2017ohk,Pietroni:2018ebj},  
here we develop a Lagrangian-coordinates approach for 
Vlasov--Poisson and detect so far unknown singularities.
These singularities comprise the intrinsic reason why 
standard perturbative techniques based on Taylor expansions 
break down at shell-crossing, while density singularities 
could be circumvented using suitable (Lagrangian) coordinates.
For simplicity, we assume a 1D (spatial) Universe;
the corresponding solutions play an important role in 3D cosmology,
mainly because shell-crossings generically begin as almost 1D phenomena
with the formation of pancakes~\citep[e.g.][]{MS:1989}.
Notwithstanding, our theoretical tools are 
scalable to any dimensions with only mild modifications.

\section{Set-up}

We denote by $q \mapsto x(q,\tau)$ the
Lagrangian map from initial \mbox{($\tau\!=\!0$)} position~$q$ to current
position ${x}$ at time~$\tau$, where~$\tau$ is not the cosmic time~$t$
but is proportional to $t^{2/3}$.  
For simplicity, we assume a spatially flat and matter dominated 
Einstein--de Sitter (EdS) universe. We make use of comoving
coordinates $x=r/a$, where $r$ is the proper space coordinate and~$a$
the cosmic scale factor (for EdS: $a =\tau$).  
The velocity is expressed in terms of the convective time derivative of
the map, i.e.,  ${v}(\xofqtext) = \partial_\tau {x}({q},\tau) =: \dot{{x}}({q},\tau)$.
The Vlasov--Poisson equations for perfectly cold dark matter are 
\begin{align} \label{EOM} 
   \ddot x  + \frac{3}{2\tau} \dot x  = - \frac{3}{2\tau} \nabla_x \varphi \,,
    \qquad \nabla_x^2 \varphi = \frac{\delta(\xofqtext)}\tau \,, 
\end{align}
where $\delta := (\rho - \bar \rho)/\bar \rho$ is the dimensionless density contrast.
Note that $\dot x =v = u/\partial_t a$ (where $u$ is the standard peculiar velocity)
has units of lengths due to the use of a dimensionless time variable~$\tau =a$; 
for simplicity, since no physical scales are introduced in an EdS universe,
we set from here on all units to unity.

Both numerical N-body methods and theory aim to solve  equations~\eqref{EOM},
however with at least one substantial difference, 
namely that in N-body methods the density contrast is  
determined by using an N-particle approximation.
By contrast, in theory one can determine the density using \citep{Taylor:1996ne}
\begin{align} \label{eq:density}
  \delta \big( \xofqtext \big) =  \int \delta_{\rm D} \big[ x(q,\tau) - x(q',\tau) \big]\, \dd q' -1  \,,
\end{align}
where ``$\delta_{\rm D}$'' is the Dirac-delta.

Observe that equations~\eqref{EOM} are invariant under the non-Galilean
coordinate transformation $x \to x + \zeta_0(\tau)$, where $\zeta_0$
is an arbitrary function of time \citep{Heckmann1955,EhlersBuchert97}. 
In the present context we use this symmetry to
 {\it enforce the following center-of-mass condition} for
the Lagrangian displacement field~$\xi(q,\tau) := x(q,\tau) -q$ on the torus $\mathbb{T}$,
\begin{align}\label{gauge} 
   \int_{\mathbb{T}} \xi(q',\tau) \,\dd q' = 0\,, \qquad
       \forall \,\, \tau > 0 \,\,.
\end{align}
Note that periodic boundary conditions are assumed in this {\it letter}.

\section{Solution strategy and initial data}

Following a standard procedure (e.g.\ \citealt{Bernardeau:2001qr}), equations~\eqref{EOM} 
and~\eqref{eq:density} can be combined into a single equation  by first taking the 
Eulerian (or Lagrangian) divergence of the former. Converting the derivatives 
according to $(\partial_q x)\,\partial_x = \partial_q$, we obtain
\begin{align}
   \label{EVO2}
   \partial_q \RT  \xi  =  - \frac 3 2 F(\xofq) \,,
\end{align}
where $\RT\!=\!\tau^2 \partial_\tau^2+({3\tau}/2)\partial_\tau\!-3/2$ is the linear 
growth operator with eigenvalues $+1$ and $-3/2$, and 
\be \label{eq:force}
 F(\xofqtext) := (\partial_q x) \int \delta_{\rm D}[ x(q,\tau) - x(q',\tau) ] \,\dd q' -1
\ee 
is the effective multi-stream force ($F=0$ in single-stream regions).
See the supplementary material~\ref{app:detailsEq4} for derivations.
Integrating~\eqref{EVO2} in space from $0$ to $q$, we obtain our main evolution equation 
\begin{align}
   \label{EVO3}
   \RT  \left\{ \xi(q,\tau) - \xi_{\rm c}(\tau) \right\}  =  - \frac 3 2 S(\xofq) \,.
\end{align}
Here, $S( \xofqtext)\!:= \!\int_0^q F(x\text{\footnotesize $(q',\tau)$}) \,\dd q'$ is 
the integrated multi-streaming force, while
$\xi_{\rm c}(\tau)\!:=\!\xi(\text{\footnotesize $q=0$},\tau)$ is 
a space-independent integration constant which, as we show,   
{\it needs to be adjusted due to multi-streaming by virtue of~\eqref{gauge}, such that 
equations~\eqref{EOM} and~\eqref{EVO3} agree with each other.}

To solve equation~\eqref{EVO3}, we provide growing-mode initial conditions
at $\tau=0$. We specify, actually without loss of generality 
(see supplementary material~\ref{app:ICs}), 
the initial velocity to be periodic,
\begin{align}\label{IC}
   \dot\xi(q, \text{\footnotesize $\tau=0$}) = 
       - \sin q + c \sin^4 q - \frac{6c}{5} \sin^6 q =: v^{\rm (ini)} \,, 
\end{align}
where $c$ is a free parameter, and we have added a counter term $\propto \sin^6 q$ 
to avoid trivial violations of~\eqref{gauge}.
In the real Universe, the initial velocity is of random nature,
which however may be locally expanded as $v^{\rm ini} = - a q + b q^3 + c q^4 + \ldots$, 
where $a-c$ are free parameters that do not alter the nature of the reported singularities,
and we have removed a quadratic term by a Galilean transformation.
We will see shortly that such low-order truncations allow us to accurately 
determine the post-shell-crossing forces by analytical means. 
Note that the setting $c=0$, as effectively employed by~\cite{Taruya:2017ohk}, 
enforces a perfectly point-symmetric collapse which we claim however is degenerate 
in a Universe with random initial conditions.
Therefore we keep~$c$ non-zero, but assume 
for simplicity that it is sufficiently small ($c \lesssim 0.49$), 
ensuring that the location of the first shell-crossing, 
controlled by the minimum of $\partial_q v^{\rm (ini)}$, 
occurs at the origin~$q=0$ in our chosen coordinate system.

\section{Shell-crossing solution}\label{sec:ZA}

In the single-stream regime the spatial integral in $F$ simplifies as there is only a 
single root $x(q,\tau) = x(q',\tau)$ that contributes to the integral, 
yielding $1/(\partial_q x)$; thus $F=0$ and so does its integral, $S=0$. Hence, 
equation~\eqref{EVO3} reduces to
$\RT \{ \xi - \xi_{\rm c}\} = 0$. 
Furthermore, due to the absence of asymmetries in the evolution equation, we have, 
by virtue of the center-of-mass condition~\eqref{gauge}, that $\xi_{\rm c}\!=\!0$. 
Thus, the evolution equation can be solved with the initial condition~\eqref{IC}, 
and we recover the well-known Zel'dovich solution~\citep{Zeldovich:1969sb}
\begin{align}\label{ZA}
  x_\ZA (q,\tau) = q + \tau \,v^{\rm (ini)}(q)\,.
\end{align}
This solution is only valid until the time of first shell-crossing, 
denoted with $\tau_\star$, that is when the particle trajectory loses its 
single-valuedness and CDM enters into the multi-stream regime.
For topological reasons (cf.\ Fig.\,\ref{fig:plot-one}), the first 
appearance of $\partial_q x_{\ZA} =0$ marks the first shell-crossing,
which, as is well known, is accompanied with an infinite density (cf.\ equation~\ref{eq:density}): 
\be
  \delta(\xofqtextZAstar) = \frac{1}{\partial_q x_{\ZA}(q,\tau_\star)|_{q=q_\star}}-1 = \infty \,.
\ee
It is easily checked that for the considered initial conditions, the first 
shell-crossing occurs at $\tau_\star =1$ at $q = q_\star=0$, for $c \lesssim 0.49$.

\section{Post-shell-crossing dynamics}

To make progress on the analysis after shell-crossing, we introduce 
an iterative scheme for~\eqref{EVO3} in which the evolution of the 
post-shell-crossing (PSC) displacement, $\xiPSC$,
is driven by an integrated force resulting from the Zel'dovich flow, i.e., 
\begin{align}\label{EVO-pza}
  \RT \!  \left\{ \xiPSC(q,\tau) - \xi_{\rm c}(\tau) \right\}  = - \frac 3 2 S_{\ZA}(q,\tau) 
\end{align}
in the first iteration, where $S_{\ZA}(q,\tau) := \int_0^q F(x_{\ZA}(q',\tau))\, \dd q'$.
In the following we summarize the main technical steps to solve~\eqref{EVO-pza}, 
while in-depth derivations and instructions for higher-order refinements are 
respectively provided in the supplementary material~\ref{app:force} 
and~\ref{app:expansion}. We remark that our iterative scheme is related 
to the one of~\cite{Taruya:2017ohk} but there are differences; 
see the supplementary material~\ref{app:otherPSCs} for details.

\begin{figure}
 \begin{center}
    \includegraphics[width=0.42\textwidth]{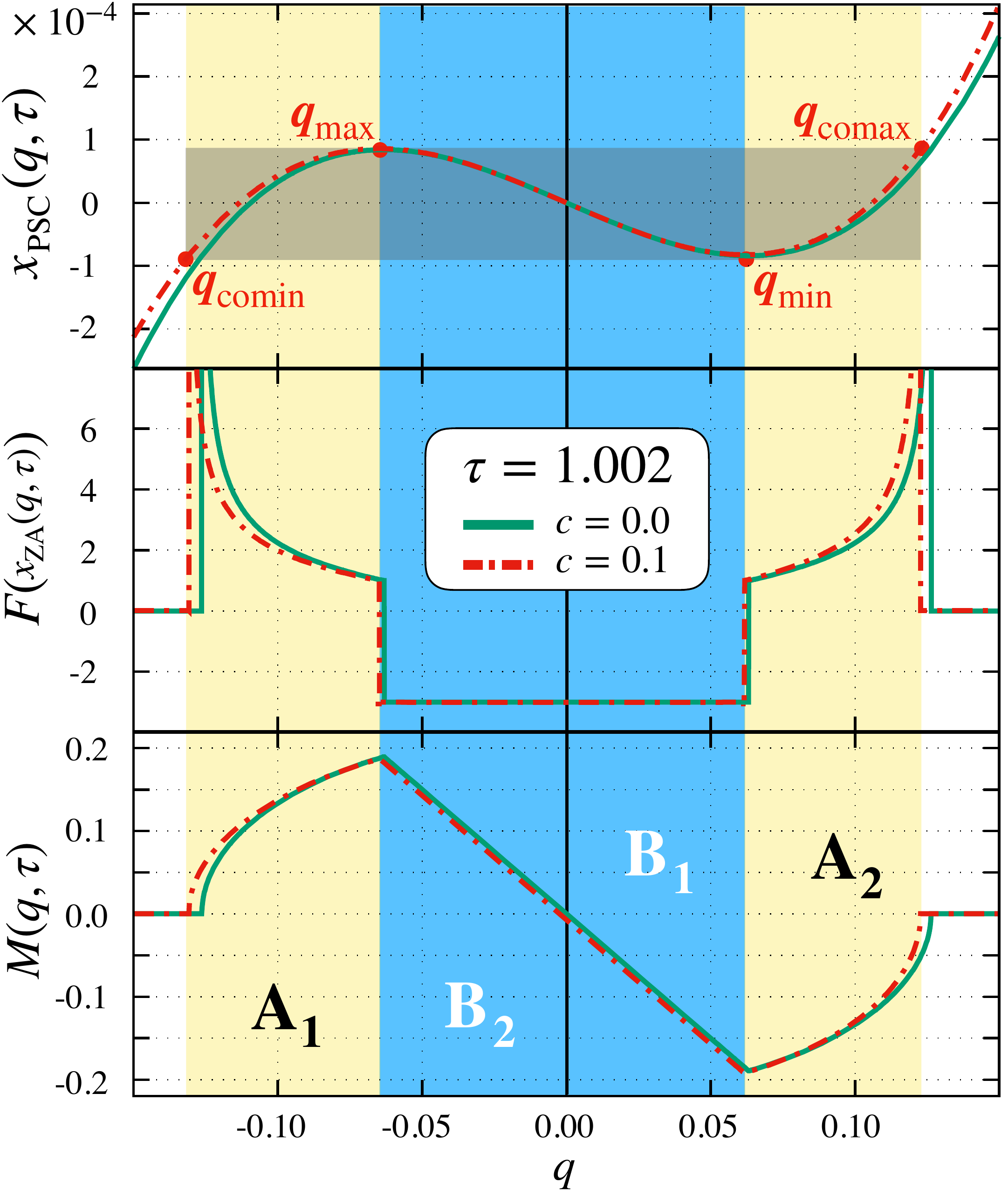}
 \end{center}
 \caption{Shown is the post-shell-crossing map $x_{\PSC}=q+\xiPSC$ (top panel), the
   multi-streaming force $F$ (middle panel) and $M := S_{\ZA} - S_{\rm c}$ (lower panel).
   All plots are evaluated after the first shell-crossing
   ($\tau_\star\!=\!1$), namely at $\tau\!=\!1.002$.  Green (solid) lines
   denote $c\!=\!0$, whereas red (dot-dashed) lines $c\!=\!0.1$.  The grey
   (horizontally) shaded region marks the multi-streaming region (for $c\!=\!0.1$), 
   which spans up the  ascending ({\bf A}$_{{\bf 1}, {\bf 2}}$;
   yellow-shaded) and the descending ({\bf B}$_{{\bf 1}, \bf 2}$; 
   blue-shaded) multi-stream regimes.
   The single-stream regime ({\bf S}$_{{\bf 1}, \bf 2}$) has no shading;  
   in this context, note that due to the assumed periodicity, 
   that $q \in [-\uppi,\uppi]$. The sharp
   non-differentiable features in~$F$ and~$M$, as well as the slight
   shift of~$F$ and~$M$ in the presence of nonzero~$c$, are
   physical effects associated to singular behavior.} 
 \label{fig:plot-one}
\end{figure}

We first need to solve for $S_{\ZA}$; for this, 
observe that the  force~\eqref{eq:force} can be written 
as~$F(\xofq)=  \int \partial_q \Theta(\xofq-\xofqprime)\, \dd q' -1$, 
where~$\Theta$ is the Heaviside step function. 
Therefore, the spatial integral of the force from 0 to $q$, within the ZA, is simply
\be \label{eq:SZAfirst}
  S_\ZA = \!\int\!\! \big[  \Theta( \xofqtextZA - \xofqtextZAprime) - \Theta(-\xofqtextZAprime) \big]  \dd q' -q ,
\ee
where we have used that $x_\ZA(\text{\small $q=0$\,}, \tau)=0$.
Thus, solving for $S_\ZA$ amounts to finding the positions (roots) when the 
arguments of the $\Theta$'s change their signs. Shortly after shell-crossing, and given
that~$c$ is small, both arguments of the $\Theta$'s in equation~\eqref{eq:SZAfirst} have 
three physical roots, implying that the flow has entered the three-stream regime 
(grey shading in Fig.\,\ref{fig:plot-one}).

Furthermore, the multi-streaming regime is still confined to small  areas around $q\!=\!0=\!q_\star$, 
thus the positions of the three roots can be obtained by considering the low-order truncation 
$ v^{\rm ini} \approx - q + q^3/6 + c\, q^4$  in $x_\ZA$ 
(higher-order terms do not change the nature of the reported singularities).
For small $c$, the positions of the three roots are slightly shifted with respect 
to the $c=0$ case, which can be determined in perturbation theory, yielding
\begin{align} \label{eq:SZA}
   S_{\ZA} \! =  S_{\rm c} +
    \begin{cases}
    \! 
    \begin{alignedat}{3}
      & \phantom{+}0; &  
          0 \!\leq\!  \tau &\!\leq\! \tau_1;  \qquad\, & \text{\small \bf S}_{{\bf 1}, \bf 2}  \! \!
      \\
      &  \!- {\rm sign}(q) \sqrt{\!D(q,\tau)};\; \,\,\qquad   &  \tau_1 \!\leq\! \tau &\!\leq\!  \tau_2; \, & \text{\small \bf A}_{{\bf 1}, \bf 2} \!\! \\
      & \!-3q -36c(1-1/\tau);   &  \tau & \!\geq\! \tau_2;  & \text{\small \bf B}_{{\bf 1}, \bf 2} \! \!
    \end{alignedat}
    \end{cases} 
\end{align}
where, stemming from the $q=0$ part of~\eqref{eq:SZAfirst},
\be
  S_{\rm c} = 36c (1-1/\tau)\, \Theta(\tau-1) \,,
\ee
to first order in $c$. Here we have defined
\be  \label{eq:disc}
  D(q,\tau)=24-3q^2-24/\tau +24 cq (3-q^2-3/\tau) \,,
\ee
as well as the space-dependent times
\be 
  \tau_{1}(q) = 8/(8-q^2-5c q^3) \,\,, \quad\,\,  \tau_{2}(q) = 2/(2-q^2-8cq^3).
\ee
Conversely, the Lagrangian positions can 
be expressed by $\tau_{1}$ and $\tau_{2}$, respectively leading
to $q_{\rm comin/comax}\!=\!\mp\!\sqrt{8(1\!-1/\tau)}-20c (1\!-1/\tau)$ 
and $q_{\rm min/max}\!=\!\pm\!\sqrt{2(1\!-1/\tau)}\!-\!8c (1\!-1/\tau)$, 
to first order in $c$. 
Also, we have $x(q_{\rm min/max})=x(q_{\rm comin/comax})$ by definition, 
where comin/comax stands for co-minimum/co-maximum.

For convenience, we have marked the positions $q_{\rm comin/comax}$ and $q_{\rm min/max}$
in Figure~\ref{fig:plot-one}, where
we also show the multi-streaming force $F(\xofqtextZA) = \partial_q S_\ZA$ and 
the integral $M := S_\ZA - S_{\rm c}$.
Observe the appearance of several non-differential features in $F(\xofqtextZA)$ and $S_\ZA$, 
indicating singularities; see the following section for a thorough analysis.  
We also note that the density is infinite at $q_{\rm min/max}$, 
where  $\partial_q x(q)|_{q=q_{\rm min/max}}=0$, which is well-known;
see e.g.\ \cite{Zeldovich:1969sb,ASZ:1982}.

Let us first derive $\xi_{\rm c}(\tau)$, which is generally nonzero due to a forcing imbalance. 
For this we solve equation~\eqref{EVO-pza} in the $q=0$ case, which, using equations~\eqref{eq:SZA} 
and~\eqref{gauge}, reduces to ($\tau \geq 1$)
\be \label{heckODE}
  \RT \xi_{\rm c}(\tau) = 54c (1-1/\tau) \,.
\ee
Supplemented with  the boundary conditions at shell-crossing 
$\xi_{\rm c}(\text{\footnotesize $\tau=1$}) =0 = \dot \xi_{\rm c}(\text{\footnotesize $\tau=1$})$,
we obtain for the particle that is initially at $q=0$ the following trajectory ($\tau \geq 1$)
\begin{align}\label{heck}
    \xi_{\rm c}(\tau) =  - \frac{18 c}{5}  \left( 10 + 8 \tau^{-3/2} -15/\tau -3 \tau \right) \,.
\end{align}
Thus, $\xi_{\rm c}$ {\it is an effective time-dependent boost that
  switches on only after shell-crossing}, which is clearly non-analytic behaviour;  
see section~\ref{sec:sing} for details.
Some indications of that boost, for  initial conditions of a distorted Gaussian 
shape, have been computed numerically in~\cite{Pietroni:2018ebj}; see their Fig.\,5.
We note that the boost~$\xi_{\rm c}$ does not only affect the $q=0$ particle, as shown below.

\begin{figure*}
 \begin{center}
    \includegraphics[width=0.95\textwidth]{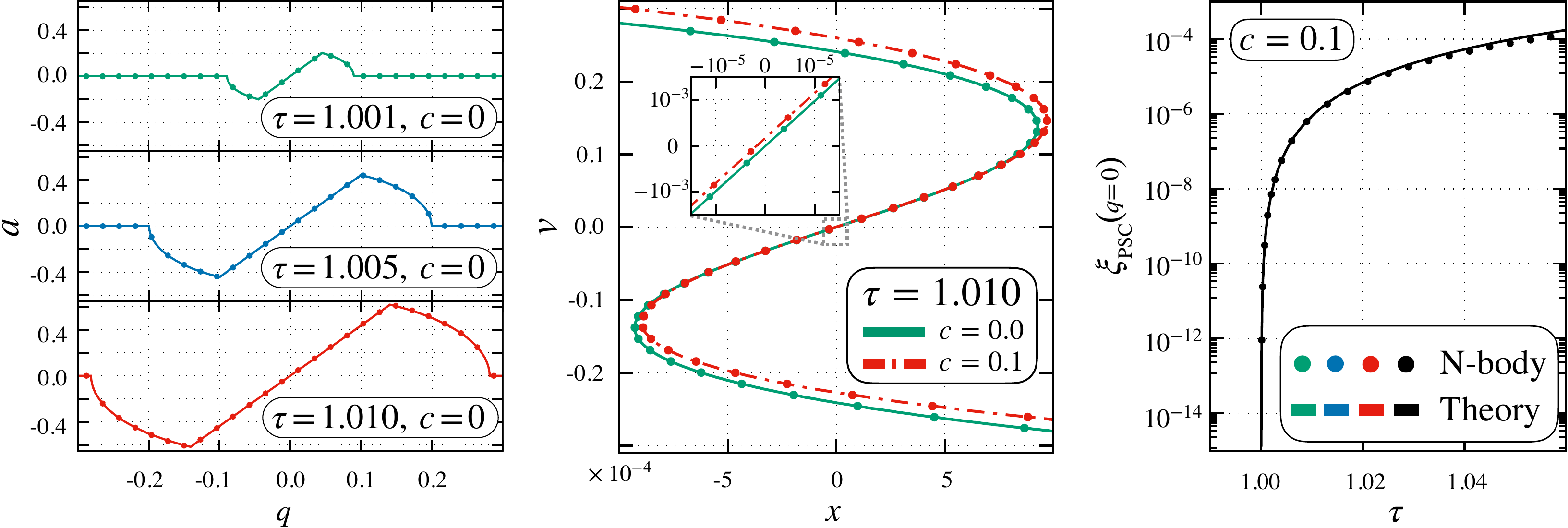}
 \end{center}
 \caption{Results of theory (solid or dot-dashed lines) against
   numerical simulations (dotted; only every 30th data
   point shown) shortly after the first shell-crossing ($\tau=1$). The
   left panel shows the acceleration $a := \ddot x_{\PSC} =
   \ddot\xi_{\PSC}$ as a function of the initial position~$q$
   for~$c=0$, displaying four non-differentiable sharp features and
   thereby unveiling singular dynamics (theory and numerics agree
   almost perfectly). The central panel shows the phase-space
   shortly after shell-crossing, for $c=0$ (green) and $c=0.1$ (red); it involves
   a slight shift of the particles in the center of the multi-stream
   regime for $c\neq0$ when the force is asymmetrical.
   The rightmost panel shows the same  effect as
   a function of time, for the particle that was initially at
   $q=0$. The results depicted here cannot be predicted by the ZA; 
   see the supplementary material~\ref{app:late-time} for details, 
   including comparisons between ZA and PSC theory as well as late-time results.  
 }
 \label{fig:plot-two}
\end{figure*}

Having obtained  $S_\ZA$ and $\xi_{\rm c}$, equation~\eqref{EVO-pza} can 
be straightforwardly solved by the method of variation of constants, 
$\xiPSC = \lambda(\tau)\, \tau + \mu(\tau) \,\tau^{-3/2}$, with 
boundary conditions provided at shell-crossing, i.e., 
$\xiPSC(\text{\footnotesize $\tau=1$}) = \dot\xi_{\PSC}(\text{\footnotesize $\tau=1$}) = v^{\rm (ini)}$, 
where $v^{\rm (ini)}$ is given in equation~\eqref{IC}.
See the supplementary material~\ref{app:xiPSC} for calculational details. We finally obtain
the post-shell-crossing displacement
\begin{subequations} \label{eqs:LPZAall}
\begin{align}\label{eq:PZAdispl}
  \xiPSC(q,\tau) = 
      \tau v^{\rm (ini)}(q) +\!
    \begin{cases}
    \! 
    \begin{alignedat}{3}
      &  0 \,; &  
          0 \!\leq\!  \tau &\!\leq\! \tau_{1};   & \text{\small \bf S}_{{\bf 1}, \bf 2}
      \\
      & \delta \xi_{\rm \bf A}  \,; \quad   &  \tau_{1} \!\leq\! \tau &\!\leq\!  \tau_{2};  \qquad & \text{\small \bf A}_{{\bf 1}, \bf 2}  \\
      & \delta \xi_{\rm \bf B}  \,;   &  \tau & \!\geq\! \tau_{2};  & \text{\small \bf B}_{{\bf 1}, \bf 2} 
    \end{alignedat}
    \end{cases}\!\!
\end{align}
where 
\begin{align}
  &\delta \xi_{\rm \bf A} =   \frac{{\rm sign}(q)}{180} \frac{D^{5/2}(q,\tau)\,\tau}{8-q^2+c q (48-11 q^2)} \,, \\
  &\delta \xi_{\rm \bf B} = \xi_{\rm c}(\tau) -3 q+ \frac{9q\tau}{20}  (4-q^2) + \frac{48}{5} \sqrt{\frac{2}{2-q^2}} \frac{q\tau^{-3/2}}{8-q^2}  \nonumber \\ 
 &\,\,  + \frac{9c}{20} \bigg[ 64\tau^{-3/2} - \tau q^4 - \frac{16q^4(3q^2-4)\tau^{-3/2} }{(1-q^2/2)^{3/2}(q^2-8)^2} 
     - 96 \tau_2^{1/2}/\tau^{3/2}    \nonumber \\
 &\,\,+ 32 \tau_2^{3/2}/\tau^{3/2} -24 \tau (1-1/\tau_2)^2 \bigg] \,.
\end{align}
\end{subequations}
The resulting map $x_\PSC(q,\tau) = q + \xiPSC(q,\tau)$ is shown in the top panel of Fig.\,\ref{fig:plot-one}. 
For convenience, in that figure
we have shaded the ascending 
({\bf A}$_{{\bf 1}, \bf 2}$, yellow)  
and descending 
({\bf B}$_{{\bf 1}, \bf 2}$, blue)  
multi-stream branches, 
while single-stream regions ({\bf S}$_{{\bf 1}, \bf 2}$) have no shading.

\section{Singularities in space and time}\label{sec:sing}

Starting from the instant of first shell-crossing, some CDM particles are 
directly exposed to infinite densities at their current positions.
We show that the displacements of those exposed particles
cannot be represented by convergent spatial or temporal 
Taylor series, 
which directly implies that standard perturbative techniques break down.
For this we identify the rational singularity exponents $\lambda$, 
which indicate the integer perturbation order $n := \lceil \lambda \rceil$ 
at which the $n$th derivative of the displacement, i.e., 
its Taylor coefficient, blows up.

Singularities in the displacement~\eqref{eqs:LPZAall} could
originate from two distinct sources, either by~(a) explicit
non-analytic features within the piecewise defined branches of the
map, and/or by~(b) discontinuities that arise when ``gluing'' the
branches together. Related to~(a), it is easily checked that the only
non-analyticity within the branches arises when $D(q,\tau)$ 
vanishes (equation~\ref{eq:disc}), which occurs at~$\tau =\tau_{1}$. 
Related to (b), two other singularities are revealed at $\tau_{2}$.

To identify the (a)-type singularity it suffices to limit ourselves to $c=0$.
Indeed, a vanishing discriminant $D$ is achieved by freezing the space
dependence in $D$ and investigating small discrepancies $\delta \tau>0$
around $\tau_{1}$ from above. 
Taylor-expanding the term with the
discriminant around $\delta \tau$ then leads to 
a $\delta \tau^{5/2}$ singularity. 
Thus, the displacement behaves locally as $\propto ( \tau - \tau_1)^{5/2}$, 
implying that its third time derivative blows up at~$\tau_1$.
Similarly, by freezing the time and
varying $q$, one finds spatial singularities at $q_{\rm comax/comin}$
with exponent $5/2$. Physically, what happens is that a spectator
particle near $q_{\rm comax/comin}$ crosses a caustic at the
current position.

The (b)-type singularities are expected to occur at $\tau_{2}$.
One of those singularities stems from glueing the ascending and 
descending multi-stream branches together, for which purpose it suffices 
to set $c=0$. Taylor-expanding around small temporal discrepancies near 
$\tau_{2}$, we find that the third-order time derivative flips sign,
indicating that the third derivative of $\xi$ is
discontinuous, thereby marking a singularity of 
$\xi \sim (\tau- \tau_{2})^3 \,\theta \left( \tau_{2} - \tau \right)$.
Similarly, we find that the third spatial derivative of the map
is also discontinuous, thus implying spatial singularities near 
$q_{\rm min/max}$ of exponent~$3$.

Lastly, a nontrivial singularity originates from $\xi_{\rm c}$
(equation~\ref{heck}), which is driven by a force imbalance (from
$c\!\neq\!0$ in equation~\ref{IC}); this manifests itself
through the loss of analyticity at the first shell-crossing, 
thereby resulting in a 
{\it dynamical phase transition due to a broken symmetry 
in the multi-streaming force.}
Indeed, the boost $\xi_{\rm c}$ ``switches'' from off to 
on ($\propto \delta \tau^3$)
once multi-streaming develops~(right panel in Fig.\,\ref{fig:plot-two}). 
Consequently, a particle that
is initially at $q=0$ will remain there until the first shell-crossing at~$\tau=1$, 
but then
the first time-derivative of its acceleration receives a non-analytic
contribution, jumping from $0$ to a finite number.

The (a)-type and first (b)-type singularities are shown in the 
first panel of Fig.\,\ref{fig:plot-two} ($c\!=\!0$), 
while the second (b)-type singularity is displayed 
in the third panel, for $c\!=\!0.1$.
{\it We confront these findings against 1D numerical (N-body) simulations} that determine 
the  exact particle forces using an efficient sorting algorithm 
\citep[for a similar implementation however in a non-periodic set-up, see][]{ColombiTouma2014}. 
The results of our simulations are marked by dots in Fig.\,\ref{fig:plot-two}.
The respective 1D code, which we make publicly 
available,\footnote{\url{https://bitbucket.org/ohahn/cosmo_sim_1d}} 
solves equations~\eqref{EOM} with a symplectic time-integrating scheme.
Initial conditions are provided by the ZA using~\eqref{IC} in a 
periodic box for $x,q \in [-\uppi,\uppi]$ 
(physical units can be trivially restored in the code if needed).
Runs were performed with $10^4$ CDM particles and time-steps, 
though for detecting the singularity stemming 
from~$\xi_{\rm c}$, temporal and spatial resolutions of up to 
$10^5$ have been used.

\section{Conclusions}

Associated to the known shell-crossing singularities 
in the cold-dark-matter density, we have identified 
three non-differentiable acceleration features. 
These features imply non-analytic behavior as 
derivatives in the phase-space blow up;
therefore, any analytical technique 
that solves for the CDM dynamics by
using temporal or spatial perturbative 
expansions will break down at shell-crossing.

Two of those singularities are of local origin and appear either when
particles enter a multi-streaming region, or when particles are
directly involved in the density caustic (Fig.\,\ref{fig:plot-one} and
first panel in Fig.\,\ref{fig:plot-two}).  Yet, a third singularity is of
global origin 
emerging from a boost in the theory, 
$\xi_{\rm c}(\tau)$, 
needed to establish momentum conservation in asymmetric collapse scenarios
(second and third panel in Fig.\,\ref{fig:plot-two}). 
In numerical simulations, by
contrast, $\xi_{\rm c}(\tau)$ appears implicitly by preserving the
fundamental conservation laws for Vlasov--Poisson.

Why are the reported singularities relatively so weak? 
Vlasov--Poisson is intimately linked to a Hamiltonian 
formulation and is therefore constrained by Liouville's theorem 
(phase-space incompressibility), which guarantees that
perfectly cold dark matter is devoid of any disruptions in the position and
velocity space (central panel in Fig.\,\ref{fig:plot-two}).  One can show that
in the generic case of initial data, which do not possess parity invariance
($c\neq0$ in equation~\ref{IC}), any attempt to construct solutions that ignore
$\xi_{\rm c}(\tau)$ will lead to severe disruptions of the phase-space,
thereby violating phase-space incompressibility.

A straightforward yet challenging extension to our work is to
exploit the  singularity theory in quasi-1D, where
departures from 1D are perturbatively small, thus providing a
bookkeeping parameter (cf.\,\citealt{Rampf:2017jan}).  
Generalizations to 3D are feasible as well, by using higher-order 
LPT~\citep{Zheligovsky:2013eca,Rampf:2015mza,Matsubara:2015ipa,2021MNRAS.501L..71R} 
and providing boundary conditions at shell-crossing,
especially for trigonometric initial
conditions, where fast Fourier transforms can be avoided 
(cf.\,\citealt{Saga:2018nud}). In the present work, 
we have provided the stepping stones for such avenues; 
indeed, equations~\eqref{EOM}--\eqref{gauge} are trivially 
generalized to arbitrary dimensions.

A full-fledged theory for the large-scale structure has the potential to advance
its theoretical predictions.  
For example, the theory can determine inputs for heavily used
effective theories of the large-scale  structure~\citep[e.g.][]{Baumann:2010tm,Porto:2013qua}.
Indeed, such effective theories incorporate shell-crossing and
multi-streaming effects through counter terms 
(with {\it a priori} unknown time dependence), 
which are usually estimated from N-body simulations.

Finally, the Vlasov--Poisson equation applies also when the gravitational
field gets replaced by an electrostatic (repulsive) field. One direct
application of our theory is the  bump-on-tail instability
in which a beam of charged particles moves in a background neutral
plasma~\citep[e.g.][]{doi:10.1063/1.1693587,PhysRevE.87.031101,Escande2018}. 
Indeed, apart from minor adaptions (e.g., sign change of charge),
our evolution equations are identical with the one of~\cite{doi:10.1063/1.1693587} 
in the continuous limit. Generalizations to multiple cold beams or to the warm 
case are straightforward too (see e.g.\,\citealt{carlevaro_falessi_montani_zonca_2015}), 
and could provide significant insights to the development of instabilities in plasma physics.

\section*{Acknowledgements}

We thank Patrick Diamond, Massimo Pietroni, Zachary Slepian and Matias Zaldarriaga for useful discussions. 
CR acknowledges funding from the People Programme (Marie Sk\l odowska--Curie Actions) of 
the European Union's Horizon 2020 Programme under Grant Agreement No.\,795707 (COSMO-BLOW-UP).
UF acknowledges financial support from the Universit\'e de la C\^ote d'Azur 
under Grant Agreement No.\,ANR-15-IDEX-01 (2018-2019). OH acknowledges funding from 
the European Research Council (ERC) under the European Union's Horizon 2020 research and 
innovation programme (Grant Agreement No.\,679145, project 'COSMO-SIMS').

\section*{Data Availability}

The code to perform N-body simulations in 1D is freely available at \url{https://bitbucket.org/ohahn/cosmo_sim_1d}. 
The data underlying this article will be shared on reasonable request to the corresponding author.

\bibliographystyle{mnras}
\bibliography{desing-v2}

\section*{Supporting information}

Supplementary data are available at {\it MNRASL} online.\\
{\bf Figure \ref{fig:heckcounter}.} The boost $\xi_{\rm c}(\tau)$ for different counter terms. \\
{\bf Figure \ref{fig:heavi}.} Analysis of the involved integrands in $S_\ZA$. \\ 
{\bf Figure \ref{fig:funnel}.} The multi-streaming force at the PSC refinement level. \\
{\bf Figure \ref{fig:latetime1}.} Late-time evolution of the boost $\xi_{\rm c}(\tau)$. \\
{\bf Figure \ref{fig:latetime2}.} Late-time evolution of the phase-space with $c=0.1$. \\
{\bf Figure \ref{fig:latetime3}.} Late-time evolution of the phase-space with $c=0.0$. \\
Please note: Oxford University Press is not responsible for the content or functionality of any supporting materials supplied by the authors.
Any queries (other than missing material) should be directed to the corresponding author for the article.

\bsp

\newpage

\appendix

\section*{{\it Supplementary material}}

\vskip0.3cm

\renewcommand\thefigure{S\arabic{figure}}    
\setcounter{figure}{0}

\section{Evolution equations and choice of initial data}

In the following we address both the aspects of our chosen theoretical starting point 
(section~\ref{app:detailsEq4}) as well as outline the arguments that support the 
generality of findings~(section~\ref{app:ICs}).

\subsection{Derivation of main evolution equation}\label{app:detailsEq4}

Here we provide details to the derivation of our main evolution equation~\eqref{EVO2}.
We begin with the standard Vlasov--Poisson equations for dark matter, which we repeat 
here for convenience (see e.g.\ \citealt{Uhlemann:2018gzz})
\begin{subequations}
\begin{align} 
  & \ddot x  + \frac{3}{2\tau} \dot x  = - \frac{3}{2\tau} \nabla_x \varphi \,, \label{eq:VP} \\
 & \nabla_x^2 \varphi = \frac{\delta(\xofqtext)}\tau   \,, \label{eq:Poiss}
\end{align}
where
\be
  \delta \big( \xofqtext \big) =  \int \delta_{\rm D} \big[ x(q,\tau) - x(q',\tau) \big]\, \dd q' -1  \,. \label{eq:delta}
\ee
\end{subequations}
To proceed, we take the Eulerian divergence of equation~\eqref{eq:VP}, which allows 
us to express its right-hand side
in terms of the Poisson equation,  leading to
\be
  \nabla_x \left( \ddot x  + \frac{3}{2\tau} \dot x\right)  = - \frac{3}{2\tau^2}  \left[ \int \delta_{\rm D} \big[ x(q,\tau) - x(q',\tau) \big]\, \dd q' -1 \right] \,.
\ee
Multiplying the last equation by $\tau^2 \partial_q x$ and
converting the Eulerian derivative according to $(\partial_q x)\,\nabla_x = \partial_q$, we obtain
\be \label{eq:almost}
  \partial_q \left( \tau^2 \ddot x  + \frac{3\tau}{2} \dot x  \right) = -  \frac{3}{2} F(\xofqtext)  + \frac{3}{2} \left(  \partial_q x - 1 \right) \,,
\ee
where we have identified the effective multi-streaming force
\be \label{eq:forceapp}
  F(\xofqtext) = -1 + (\partial_q x) \int \delta_{\rm D} \big[ x(q,\tau) - x(q',\tau) \big]\, \dd q'  \,.
\ee
Finally, we express in equation~\eqref{eq:almost} the Lagrangian map in terms of the displacement~$\xi = x-q$, which leads to our main evolution equation
\be \label{eq:mainRep}
  \partial_q \RT \xi = - \frac 3 2 F \,,
\ee
where $\RT= \tau^2 \partial_\tau^2+({3\tau}/2)\partial_\tau\!-3/2$. We note that the Lagrangian space derivative commutes with the Lagrangian time derivative.

One particularly convenient feature of equation~\eqref{eq:mainRep} is that, in single-stream regions for which $F=0$, it trivially integrates to the standard differential equation for the Zel'dovich displacement (cf.\ section~\ref{sec:ZA} in the main text).
This feature allows to evolve all particles in single-stream regions using the Zel'dovich displacement, as well as providing accurate boundary conditions once particles are about to enter multi-streaming regions.

As regards to the computations of the multi-streaming force, equation~\eqref{eq:forceapp} involves a spatial integral over a Dirac delta which, upon spatial integration, can be 
converted into a Heaviside step-function $\Theta$.
 As we elucidate in detail in App.\ \ref{app:force}, determining the integrated multi-streaming force using $\Theta$'s  amounts to a significant simplification.
This is important for the present considerations, since integrations can be nontrivial due to the non-analytic nature of the force in multi-streaming regions.

\subsection{Initial conditions and generality of findings} \label{app:ICs}

In the present paper we choose for the initial peculiar velocity
\be \label{eq:vini}
  v^{\rm (ini)} =  - \sin q + c \sin^4 q - \frac{6c}{5} \sin^6 q  \,,
\ee
where we assume that $c$ is a small parameter. Although this initial velocity appears to be quite specific, we show in the following that our choice of initial data is actually generic, in the sense that the nature of the reported singularities does not change for similar choices of initial data (modulus trivial changes in prefactors etc.).

First of all, we use smooth periodic initial data as they are in accordance with the known observational fact of the random nature of primordial density fluctuations (with vanishing mean). 
We remark that  non-smooth initial conditions (such as logarithms) would imply singular initial conditions, whereas it is our goal here to see whether {\it smooth} initial conditions lead eventually to {\it singularities.}

At the technical level for detecting the singularities, we need to determine the particle forces shortly after shell-crossing to sufficient accuracy. For this it is crucial to 
notice that the spatial width of the multi-streaming region is very narrow shortly after shell-crossing,
 implying that low-order spatial truncations of the {\it initial data} are fully meaningful  
(in catastrophe theory, this is called a ’normal form’ approach). 
Thus, to determine the force shortly after shell-crossing, it suffices to approximate~\eqref{eq:vini} with the spatial expansion
\be \label{eq:vininormal}
   v^{\rm (ini)} = -q + \frac{q^3}{3!} + cq^4 + H.O.T., 
\ee
where $H.O.T.$ stands for higher-order terms; see Appendix~\ref{app:expansion} for higher-order refinements.
We remark here that we exclude terms $\sim q^2$ in~\eqref{eq:vininormal} (e.g., stemming from a cosine in equation~\ref{eq:vini}); indeed, their effect is boosting the system by a constant velocity, which can be removed by a Galilean transformation. 
Therefore, while the precise time and location of the first shell-crossing might change with such a new term, no new physics is added and thus, we can safely ignore such terms.

Our choice of initial data is fully generic also in the sense that one could decorate various numerical coefficients in front of the spatial terms appearing in the normal form~\eqref{eq:vininormal}, which would not alter the nature of the reported singularities (but could clutter the analytical expressions).

\begin{figure}
 \begin{center}
    \includegraphics[width=0.46\textwidth]{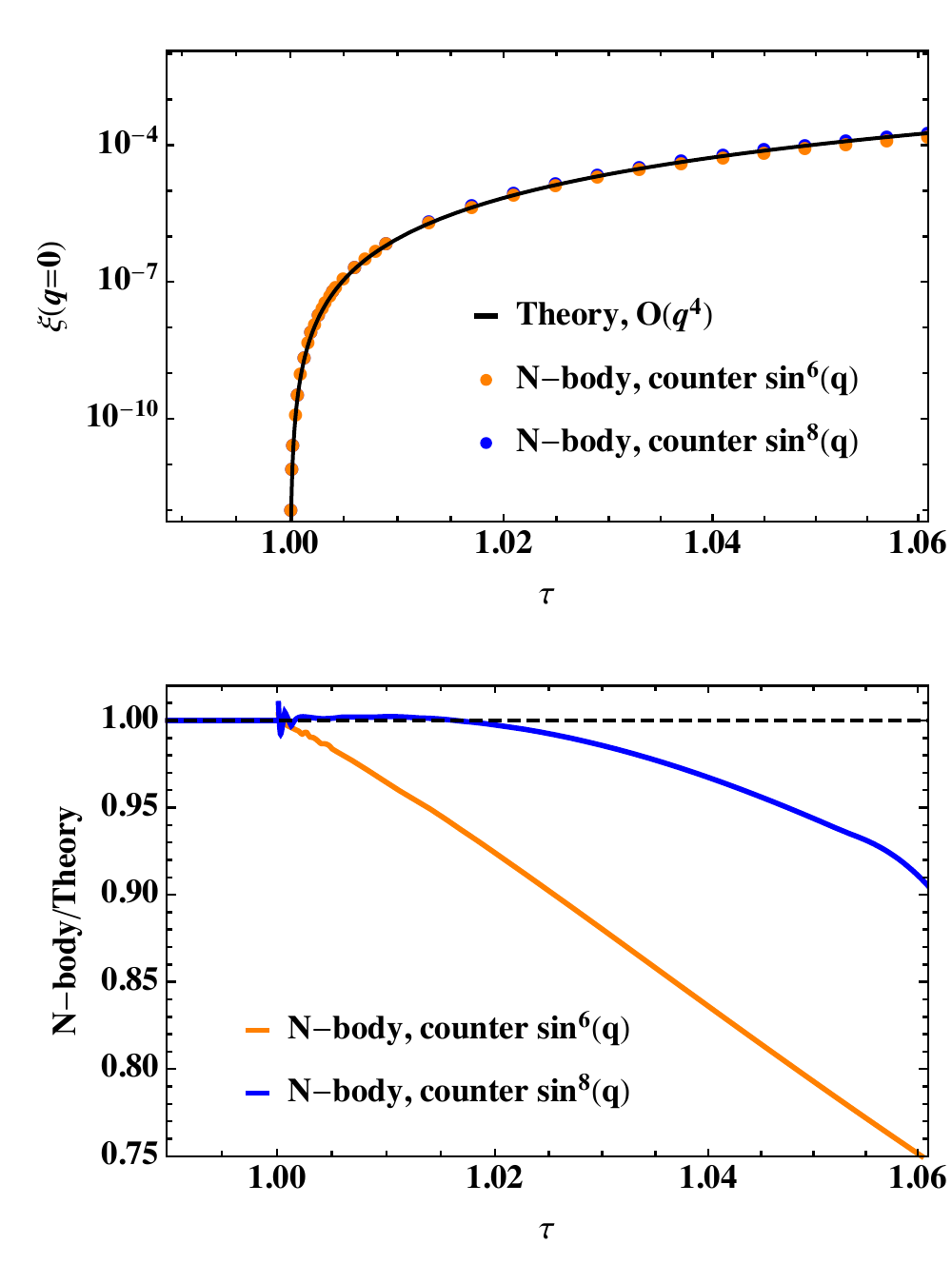} 
 \end{center}
\vskip-0.2cm
 \caption{The late time-effect of adding higher-order counter terms in the initial data on the boost  $\xi_{\rm c}(\tau)= \xi(q=0,\tau)$, for $c=0.1$.
   {\bf Top:} 
   The solid line denotes our theory results, while the orange [blue] dots denote simulation results using equation~\eqref{eq:vini} [equation~\eqref{eq:vinitilde}] for the initial data.
   {\bf Bottom:} Same as above but shown is the ratio N-body/theory. We note that the deviations in the ratio shortly after $\tau= \tau_\star=1$ are due to limited resolution in the numerical simulation.} 
 \label{fig:heckcounter}
\end{figure}

Finally, the nature of the singularity related to the boost is unchanged, even if one employs different counter-terms in the initial velocity. This is illustrated in Fig.\,\ref{fig:heckcounter}, where we show the particle trajectory at $q=0$ at times after shell-crossing, comparing the effects from different initial data while setting $c=0.1$.
Specifically, the red dots and black line display results from using~\eqref{eq:vini} as the initial data, while for the blue dotes  we use 
\be \label{eq:vinitilde}  
  \tilde v^{\rm (ini)} =  - \sin q + c \sin^4 q - \frac{48c}{35} \sin^8 q
\ee 
instead. As seen in   Fig.\,\ref{fig:heckcounter}, until shortly after shell-crossing, the depicted trajectory for the two sets of initial data coincide to good accuracy,
especially for the initial data that includes the counter term $\sim c q^8$.
Only at much later times the trajectories begin to differ, which is expected from theory grounds: at those late times the multi-stream regime has substantially gained on spatial width, where the  counter terms in the initial data have increasing impact.

\section{Multi-streaming force and post-shell-crossing displacement}\label{app:force}

Here we provide calculational details about how the multi-streaming force~$F$ (defined around equation~\ref{EVO2}), its spatial integral $S = \int F(q') \dd q'$ and the resulting post-shell-crossing displacement $\xi_\PSC$ are determined within the first PSC iteration, i.e., when the force, shortly after shell-crossing, is estimated using the Zel'dovich approximation.

We begin with the definition of the ZA force for which we approximate $x(q,\tau) \simeq x_\ZA(q,\tau)$ in $F$ (this approximation is exact until shell-crossing and approximative shortly after), i.e., 
\be \label{eq:ZAforce}
 F(\xofqtextZA) =  (\partial_q x_\ZA) \int \delta_{\rm D} \big[ x_\ZA(q,\tau) - x_\ZA(q',\tau) \big]\, \dd q' -1  \,,
\ee
where  $x_\ZA(q,\tau) = q +\tau \,v^{\rm (ini)}$,
with
\be \label{eq:ICrep}
 v^{\rm (ini)} = - \sin q + c \sin^4 q - \frac{6c}{5} \sin^6 q \,.
\ee
From equation~\eqref{eq:ZAforce} it is clear that one could evaluate the density locally, namely by exploiting the composition property of the Dirac-delta, that for any function $x(q)$, we have 
$\delta_{\rm D}( x(q)) = \sum_{i} \delta_{\rm D}(q-q_i)/ | \partial_q x(q_i)|$, where $q_i$ is the $i$th root of $x(q)$. 

Alternatively, and as done in the main text, we can write equation~\eqref{eq:ZAforce} equivalently as
\be
  F(\xofqtextZA) =  \int \partial_q \Theta \big[ x_\ZA(q,\tau) - x_\ZA(q',\tau) \big]\, \dd q' -1  \,,
\ee
which can be trivially integrated to $S_\ZA = \int_0^q F(\xofqtextZAprime)\, \dd q'$, yielding
\be \label{eq:SZArepApp}
  S_\ZA = \!\int\!\! \big[  \Theta( \xofqtextZA - \xofqtextZAprime) - \Theta(-\xofqtextZAprime) \big]  \dd q' -q \,.
\ee
Using the step-function, $\Theta$, instead of the Dirac-delta simplifies the calculations significantly, especially when determining the boost $\xi_{\rm c}$. Indeed, the boost arises from global constraints related to the center-of-mass condition~\eqref{gauge}, which can be easily incorporated in~\eqref{eq:SZArepApp} by integration over the torus. By contrast, solving for the boost  $\xi_{\rm c}$  in the local ``Dirac-delta approach'' is non-trivial: it can be done by considering the spatial average of $\xi_\PSC$ or, equivalently, by a  weighted spatial average of its evolution equation.
Nonetheless, we have explicitly verified that our final results coincide, whether one uses the Dirac-delta or the Heaviside approach.

In App.\,\ref{app:forcenoc} we derive directly $S_{\rm ZA}$ by solving~\eqref{eq:SZArepApp},  from which one can easily obtain $F(\xofqtextZA) = \partial_q S_\ZA$. Then, in App.\,\ref{app:xiPSC} we determine the post-shell-crossing displacement.

\subsection{Solving for the multi-streaming force}\label{app:forcenoc}

It is convenient to split up equation~\eqref{eq:SZArepApp} into two distinguished terms, i.e., $S_{\rm ZA} := S_{\rm ZA,1} + S_{\rm ZA,2}$, with
\begin{subequations} \label{subeq}
\begin{align}
  &S_{\ZA, 1} = - \int  \Theta \left( -\xofqtextZAprime \right)   \dd q' \,, \label{eq:ZA1} \\
  &S_{\ZA, 2} = \int \Theta \left( \xofqtextZA - \xofqtextZAprime \right)  \dd q' -q \,. \label{eq:ZA2}
\end{align}
\end{subequations}
Before shell-crossing, solving these equations is trivial, as both Heaviside functions have only one root. Thus, in the single-stream regime, the integration over the torus leads to $S_{\rm ZA,1} = -\uppi$ and $S_{\ZA,2} = +\uppi$; thus~$S_\ZA=0$.

After the first shell-crossing, for sufficiently small $c$, which we assume throughout this work,
there are Eulerian locations with three  fluid streams (see Fig.\,\ref{fig:plot-one}),  meaning that at such locations the arguments of the Heaviside functions in~\eqref{subeq} have each three roots.
 Furthermore, 
for given initial data and for short times after shell-crossing, the three-stream regime is confined to small spatial areas around $q=0$ 
(i.e., close to the Lagrangian locations where the first shell-crossing occurred).
This implies that, to find the three roots  shortly after shell-crossing, 
it suffices to consider the low-order spatial truncation (i.e., the normal form)  of the initial data~\eqref{eq:ICrep}, which is
\be \label{eq:viniRep}
  v^{\rm (ini)} =  - q + q^3/6 + cq^4 + H.O.T. \,,
\ee
to be used in the Zel'dovich map $x_\ZA = q + \tau v^{\rm (ini)}$.  Furthermore, since~$c$
is assumed to be small, the effect of the term $cq^4$ in~\eqref{eq:viniRep} is to shift the positions of some of the roots; in the following we will determine this effect using linear perturbation theory.

\setcounter{figure}{1} 

\begin{figure}
 \begin{center}
    \includegraphics[width=0.44\textwidth]{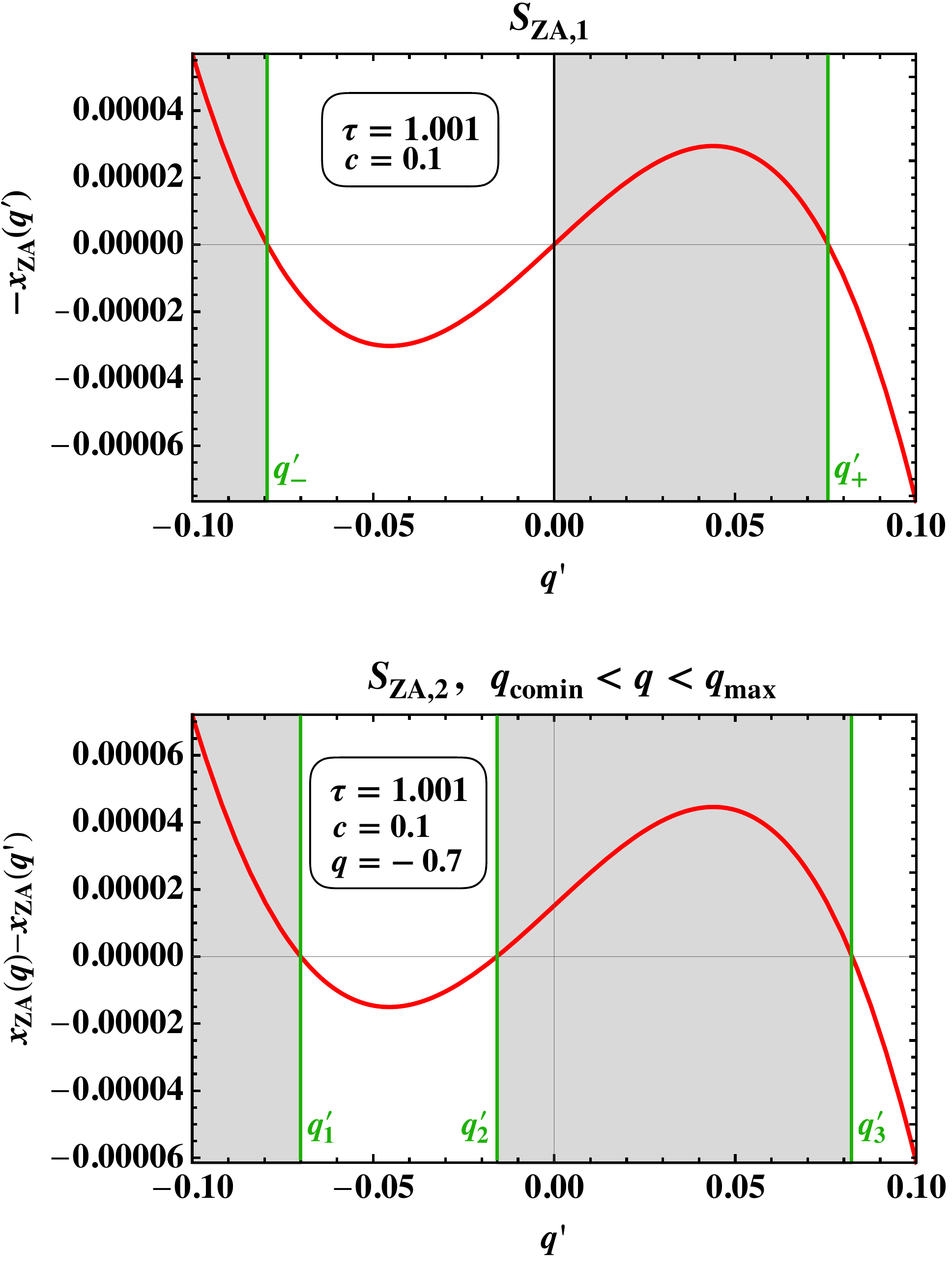}
 \end{center}
 \caption{Shown are the arguments of the Heaviside functions appearing in the spatial integrals of $S_\ZA$ according to equation~\eqref{subeq}, shortly after shell-crossing and for $c=0.1$. {\bf Top:} $S_{\ZA,1}$, which does not depend on $q$. {\bf Bottom:} $S_{\ZA, 2}$ for the case when $q_{\rm comin}< q < q_{\rm max}$. }
 \label{fig:heavi}
\end{figure}

We begin with solving for $S_{\ZA, 1}$ according to equation~\eqref{eq:ZA1}, which has the integrand $\Theta \left( -\xofqtextZAprime \right)$, with its argument shown in the upper panel of Fig.\,\ref{fig:heavi}. Shortly after shell-crossing and 
for vanishing $c$, the argument $ -\xofqtextZAprime$ has the three roots 
\be
  q' = 0\,, \qquad \bar q_\pm' = \pm \sqrt{6(1-1/\tau)} \,,
\ee
where the overbar denotes the $c=0$ case. To include the effect of $c$, we set
\be
  q_\pm' = \bar q_\pm' + \delta q_\pm \,,
\ee  
where $\delta q_\pm = O(c)$,
from which one easily finds that
\be
 q_\pm' = \pm \sqrt{6(1-1/\tau)} - 18c (1-1/\tau)  + O(c^2) \,.
\ee
These roots are also shown in the upper panel of Fig.\,\ref{fig:heavi} as vertical green lines, 
while the grey shading indicates the regime where $-\xofqtextZAprime$ is positive.
From this one then finds that
\be \label{eq:SZA1app}
 S_{\ZA,1} = -\uppi + 
     \begin{cases}
       \begin{alignedat}{2}
          & 0 \,;                  &  \tau <1  \\
          &36 c (1-1/\tau) + O(c^2)\,; \qquad  & \tau \geq 1 
       \end{alignedat}
     \end{cases} \,.
\ee

Next we consider $S_{\ZA,2}$, which involves an integral over a Heaviside function with argument
$\xofqtextZA - \xofqtextZAprime$. In contrast to the previous calculation, the integral argument depends on $q$. 
Specifically, shortly after shell-crossing, solving $x_\ZA(q,\tau) - x_\ZA(\bar q',\tau) =0$  is, to leading order,
\be \label{eq:diracargument}
  \left( q- q' \right) \left[   q'^2 + q  q'  + q^2  + 6/\tau -6  + 6 c (q + q') (q^2 + q'^2) \right] = 0 \,.
\ee
For vanishing $c$, the three roots  of this equation are
\be \label{eq:tripleroot}
  \bar q'_{1} = q\,, \qquad \bar q'_{2,3} = \frac{-q \pm \sqrt{\bar D(q,\tau)}}{2} \,,
\ee
where  $\bar D(q,\tau) = 24 -3q^2 -24/\tau$.
Similarly as above, the effect of small $c$ can be incorporated using perturbation theory, leading to
\be
   q'_{1} = q\,, \qquad q'_{2,3} = \frac 1 2 \left( -q - 36c (1-1/\tau) \pm \sqrt{D(q,\tau)} \right)
\ee
to first order in $c$, where 
\be
  D(q,\tau) := 24-3q^2 -24/\tau + 24 c q \left( 3-q^2-3/\tau \right) \,.
\ee
In the lower panel of Fig.\,\ref{fig:heavi}, we show the 
three roots $q_{1,2,3}'$ for the case when $q_{\rm comin} < q < q_{\rm max}$ which is within the ascending 
multi-stream regime ``{\bf A}${}_{\fett{1}}$'' (see Fig.\,\ref{fig:plot-one}),
where
\begin{subequations} \label{subeqQminmax}
\begin{align}
 & q_{\rm min/max} = \pm \sqrt{2(1-1/\tau)} - 8c (1-1/\tau)  + O(c^2) \,, \\
 & q_{\rm comin/comax} = \mp \sqrt{8(1-1/\tau)} - 20 c (1-1/\tau)  + O(c^2)\,.
\end{align}
\end{subequations}
The values~\eqref{subeqQminmax} can be determined as follows. 
 Firstly, $q_{\rm min/max}$ are the
locations in the multi-stream regime where the map $x_\ZA$ is minimal/maximal. Since the multi-streaming region is very narrow around $q=0$, normal-form
considerations are sufficient to determine $q_{\rm min/max}$. Demanding that $\partial_q (q  +\tau [- q + q^3/6 + cq^4]) \stackrel ! =0$, and solving it with perturbation theory then leads to the above reported $q_{\rm min/max}$.
Secondly, $q_{\rm comin/comax}$ for which $x_\ZA( q_{\rm comin/comax}) = x_\ZA( q_{\rm min/max})$ by definition, mark the locations where the flow changes from single- to multi-stream; its values can be determined by demanding that the discriminant of the reduced polynomial (i.e., the square bracketed term in equation~\ref{eq:diracargument}) vanishes.

Using this, one straightforwardly obtains
\be
   S_{\ZA, 2} = \uppi + 
     \begin{cases}
       \begin{alignedat}{2}
          & 0 \,;                &  q  <  q_{\rm comin}                  \,\,\,\land\,\,\, q >    q_{\rm comax}  \\
          &  - {\rm sign}(q) \sqrt{D(q,\tau)}\, ; \qquad   &  q_{\rm comin/min} < q <  q_{\rm max/comax}    \\
          &  - 3q - 36c (1-1/\tau)\,;     &    q_{\rm max} < q  <   q_{\rm min} 
       \end{alignedat}
     \end{cases} 
\ee
Summing up, one obtains the integrated multi-streaming force $S_\ZA = S_{\ZA, 1}+S_{\ZA, 2}$, i.e., 
\begin{align}
  &S_\ZA =  36c (1-1/\tau) \, \Theta(\tau -1)  \nonumber \\
  &\phantom{+}\qquad + \begin{cases}
    \begin{alignedat}{3}
      & \phantom{+}0; &  
          0 \!\leq\!  \tau &\!\leq\! \tau_{1};  \qquad\, & \text{\small \bf S}_{{\bf 1}, \bf 2}  \! \!
      \\
      &  \!- {\rm sign}(q) \sqrt{\!D(q,\tau)};\; \,\,\qquad   &  \tau_{1} \!\leq\! \tau &\!\leq\!  \tau_{2}; \, & \text{\small \bf A}_{{\bf 1}, \bf 2} \!\! \\
      & \!-3q - 36c (1-1/\tau);   &  \tau & \!\geq\! \tau_{2},  & \text{\small \bf B}_{{\bf 1}, \bf 2} \! \!
    \end{alignedat}
    \end{cases} \label{eqSZArepfinal}
\end{align}
to first order in $c$, where 
\be 
  \tau_{1} = 8/(8-q^2-5c q^3) \,\,, \qquad\,  \tau_{2} = 2/(2-q^2-8cq^3)
\ee
are the inverted functions of $q_{\rm comin/comax}(\tau)$ and $ q_{\rm min/max}(\tau)$, respectively.

Finally, having determined $S_\ZA$ allows us to derive $F(\xofqtextZA) = \partial_q S_\ZA$ which we report here for future reference, we find
\be \label{eq:FZA}
  F(\xofqtextZA) = 
   \begin{cases}
     \begin{alignedat}{2}
       & 0 \,;                                                        &  q  <  q_{\rm comin}                  \,\,\,\land\,\,\, q >    q_{\rm comax}  \\
       & \frac{A(q)\,{\rm sign}(q)}{\sqrt{D(q,\tau)}}  \,; \quad   &  q_{\rm comin/min} < q <  q_{\rm max/comax}   \\
       & -3 \,;                                                    &    q_{\rm max} < q  <   q_{\rm min} 
     \end{alignedat}
   \end{cases} 
\ee
to first order in $c$, where we have defined $A(q) :=3q - 36c (1-q^2-1/\tau)$.
This force is shown in the central panel of Fig.\,\ref{fig:plot-one} shortly after 
the first shell-crossing. Observe the slight shift of the force along the  horizontal 
axis in the presence of small but nonzero $c$, breaking the spatial axis-symmetric configuration. 
This symmetry breaking is precisely the driver for the boost $\xi_{\rm c}$ to emerge.

\subsection{Post-shell-crossing displacement}\label{app:xiPSC}

Once the integrated force $S_\ZA$ and the boost $\xi_{\rm c}$ are determined, 
the post-shell-crossing displacement $\xi_\PSC$ can be obtained from
\be \label{eq:ODExiPSC}
  \RT \xi_\PSC =  - \frac 3 2 S_\ZA  \,.
\ee
This is an inhomogeneous differential equation and can be solved by the method of variation of constants.
For this observe that the homogeneous equation $\RT \xi_\PSC =0$ has the general solution $\xi_\PSC = \lambda \tau + \mu \tau^{-3/2}$, where in the homogeneous
case, $\lambda$ and $\mu$ are purely spatial.
In the inhomogeneous case, these constants are allowed to depend also on time, i.e., 
\be \label{eq:ansatzPSC}
   \xi_\PSC = \lambda(\tau)\, \tau + \mu(\tau)\, \tau^{-3/2} \,,
\ee
which we supplement with the constraint 
\be  \label{constraint}
  \dot \lambda \tau + \dot \mu \tau^{-3/2} =0\,,
\ee 
together with the boundary conditions provided at the time of first shell-crossing,
\be \label{boundaries}
  \xiPSC(\text{\footnotesize $\tau=1$}) =  v^{\rm (ini)} \,, \qquad \dot\xi_{\PSC}(\text{\footnotesize $\tau=1$}) =  v^{\rm (ini)} \,,
\ee
where $v^{\rm (ini)}$ is given in equation~\eqref{IC}.
To proceed, we plug~\eqref{eq:ansatzPSC} into equation~\eqref{eq:ODExiPSC}, leading to
\be
 \tau^2 \dot \lambda - \frac 3 2 \dot \mu \tau^{-1/2} = -\frac 3 2 S_\ZA \,.
\ee
This equation together with the constraint~\eqref{constraint} form a closed set of equations for $\dot\lambda$ and $\dot\mu$; integrating 
the resulting expressions in time from 1 to $\tau$, and using the boundary conditions~\eqref{boundaries}, we find
\begin{align} 
 \begin{aligned} \label{lambdamu}
  &\lambda(\tau,q) = v^{\rm (ini)} -  \frac 3 5 \int_1^\tau \frac{S_\ZA(\eta)}{\eta^2} \dd \eta \,, \\
  &\mu(\tau,q) =  \frac 3 5 \int_1^\tau \eta^{1/2} S_\ZA(\eta) \,\dd \eta \,.
 \end{aligned}
\end{align}
Together with the explicit solution for $S_\ZA$ given in equation~\eqref{eqSZArepfinal},
it is straightforward yet tedious to integrate the expressions~\eqref{lambdamu}, which then leads 
to $\xi_\PSC$ as shown in equation~\eqref{eq:PZAdispl}. For this we remark that the first term on 
the right-hand-side of equation~\eqref{eqSZArepfinal} has no contribution to $\xi_\PSC(q)$ for $q\neq0$, 
as can be easily verified by employing the center-of-mass condition~\eqref{gauge}.

\section{Expansion scheme and higher-order force}\label{app:expansion}

In Appendix~\ref{app:expansionscheme},
we provide some  details about the employed post-shell-crossing expansion scheme. Then, in Appendix~\ref{app:forcePPSC}, 
we employ the post-shell-crossing results 
from the main text in order to determine the multi-streaming force $F$ in the next (``post-post'') iteration.

\subsection{Expansion scheme}\label{app:expansionscheme}

 For convenience let us repeat the evolution equation for the  displacement field which is
\be \label{eq:evorep}
 \partial_q \RT  \xi  =  - \frac 3 2 F(\xofq) \,,
\ee 
where $\RT\!=\!\tau^2 \partial_\tau^2+({3\tau}/2)\partial_\tau\!-3/2$, and
\be
 F(\xofqtext) =  \int \partial_q\Theta[ x(q,\tau) - x(q',\tau) ] \,\dd q' -1
\ee
is the multi-streaming force. 
To solve equation~\eqref{eq:evorep} we employ an iterative expansion scheme in which the ODE of the $n$th-order approximation of the displacement,
here denoted with $\xi_n$, is sourced by the multi-streaming force that is determined from the displacement at order $n-1$.
Thus, in our iterative expansion scheme we solve for 
\begin{subequations} \label{app:expansioschemePP}
\begin{align}
 &\partial_q \RT  \xi_\PSC  \simeq  - \frac 3 2 F_\ZA \,, \\
 &\partial_q \RT  \xi_\PPSC  \simeq  - \frac 3 2 F_\PSC \,, \label{evo:PPSC}\\
  &\partial_q \RT  \xi_\PPPSC  \simeq  - \frac 3 2 F_\PPSC \,,
\end{align}
\end{subequations}
and so on, where  ``PSC'', ``PPSC'', etc.\ denote subsequent higher-order refinements within the post-shell-crossing calculation, and we have introduced the 
shorthand notation
\be
  F_\ZA = F(\xofqtextZA) \,, \qquad  F_\PSC = F(\xofqtextPSC) \,, \qquad \text{etc.}
\ee
We remark that our expansion scheme differs from that of \cite{Taruya:2017ohk}; specifically they suggest to incorporate 
higher-order refinements by temporal and/or spatial expansions without using~\eqref{app:expansioschemePP}. More details to the different
approaches used in the literature are provided in App.~\ref{app:otherPSCs}.

\subsection{Force in the post-post-shell-crossing approximation} \label{app:forcePPSC}

Since we have already determined the displacement within the first refinement level in the main text, i.e., $\xi_\PSC$ (see equation~\ref{eq:PZAdispl}), we can use this result to determine the force for the next refinement level. Such avenues are an important consistency check: If higher-order corrections to the displacement are not remaining small, it could indicate that the proposed expansion scheme is ill-defined.

\setcounter{figure}{2} 

\begin{figure}
 \begin{center}
    \includegraphics[width=0.47\textwidth]{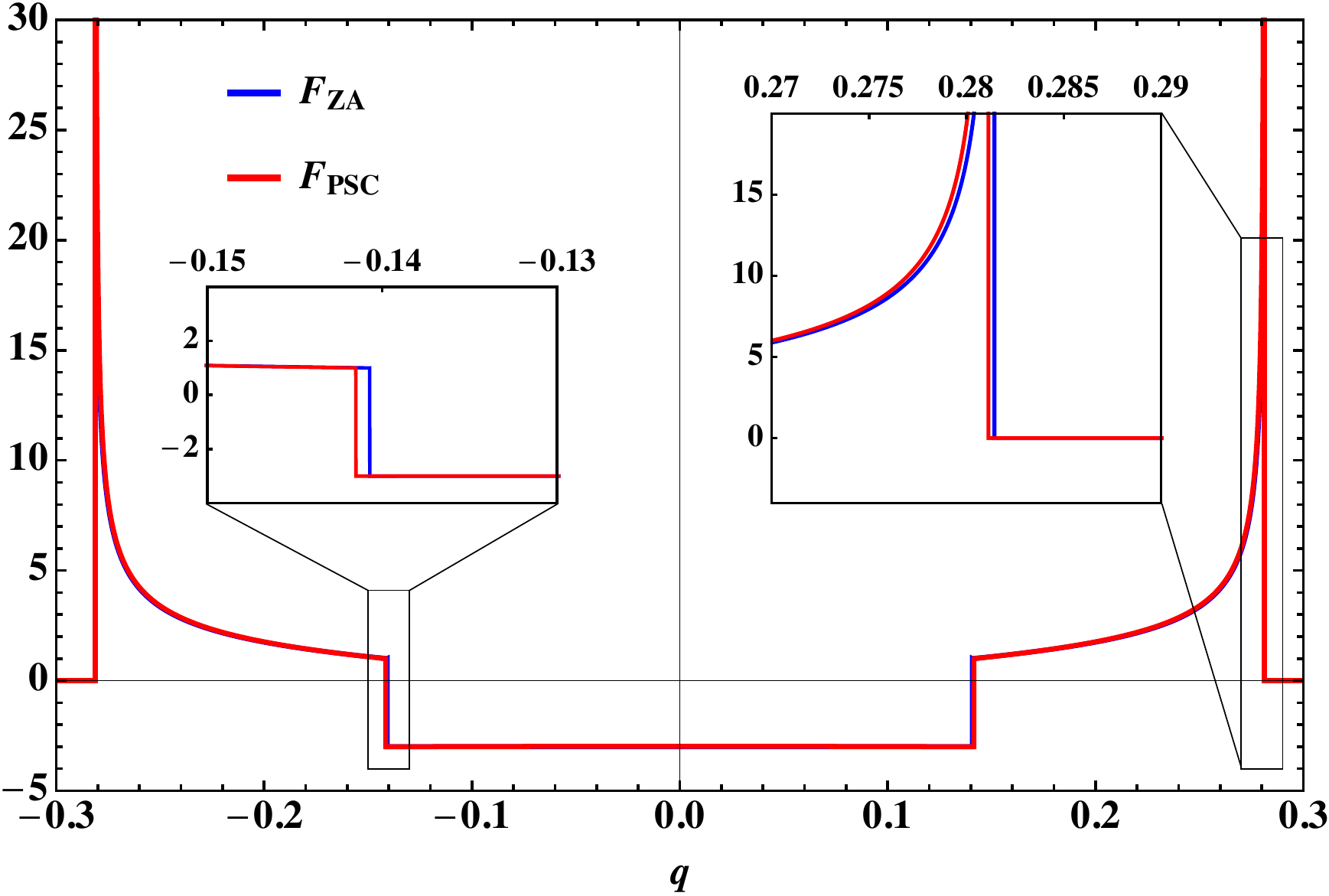}
 \end{center}
 \caption{Multi-streaming force within our expansion scheme, at $\tau=1.01$, shortly after the first shell-crossing. The blue line denotes the force within the ZA, while the red line is the updated force at the PSC refinement level. For simplicity, we consider the symmetric case  ($c=0$), i.e., we have $F_{\ZA}(q)=F_{\ZA}(-q)$ and  equivalently for the PSC force.}
 \label{fig:funnel}
\end{figure}

Specifically, in our expansion scheme the post-post-shell-crossing  displacement $\xi_\PPSC$  is governed by equation~\eqref{evo:PPSC}, which requires the multi-streaming
force $F_\PSC$ as input. In Fig.~\ref{fig:funnel} we show this updated force (red lines),  together with the Zel'dovich force (blue lines), at times shortly after shell-crossing $\tau =1.01$ for the $c=0$ case of ``symmetric'' initial data~\eqref{IC}.
The overall shape of the force is essentially unchanged when going to the higher-order refinement level, however we notice a slight shift of the pronounced features  in the force.
These shifts are due to the slight change of the spatial locations of $q_{\rm min/max}$ and $q_{\rm comin/comax}$, which can be determined  similarly as outlined in Section~\ref{app:forcenoc}. Specifically, to get $q_{\rm min/max}$ one evaluates $\partial_q  (q + \xi_\PSC) \stackrel !=0$ in perturbation theory, while to get $q_{\rm comin/comax}$ one needs to find the positions $q'$ where the discriminant of the reduced polynomial $[x_\PSC(q,\tau) - x_\PSC(q',\tau)]/(q-q')$ vanishes.
More specific details about higher-order iterations beyond shell-crossing will be assessed in future work.

\section{Comparison of analytic derivations with others in the literature}\label{app:otherPSCs}

As mentioned in the main text, there exist already a few post-collapse approaches in the literature  
\citep{Colombi:2014lda,Taruya:2017ohk,Pietroni:2018ebj}. Here we provide some details
to the differences and what the approaches have in common.

In doing so we distinguish between approaches that employ initial data without and with symmetry breaking terms (the $c$ term in our notation; cf.\ equation~\ref{IC}).
Of course, the approaches that incorporate symmetry breaking terms  have the potential to resolve the resulting non-analyticity at shell-crossing (a.k.a.\ the boost).

\paragraph*{Post-shell-crossing approaches in the $c=0$ case.}
The cosmological approach of \cite{Taruya:2017ohk}, itself based on the pioneering work of \cite{Colombi:2014lda} in a non-cosmological setup,
introduced the first analytical approach for determining the post-shell-crossing dynamics.
Similarly to our approach, \cite{Taruya:2017ohk} developed the techniques in 1D, although their approach should be scalable to 3D in a similar way as ours.

The central part of post-shell-crossing studies is determining the effective (multi-streaming) force, for which \cite{Taruya:2017ohk} solve the Poisson equation in a periodic
box using a Green's method,  taking the Ewald summation of periodic repetitions into account. 
To determine the density in this method one needs to perform a spatial integration over the Green's function times the density; the density is determined 
using the normal form of the Zel'dovich displacement (a cubic polynomial in $q$; cf.\ equation~\ref{eq:vini}, or their equation~28).
Some of the involved spatial integrals, however, need to be approximated by suitable Taylor expansions. By contrast, 
in our approach,  we determine the Zel'dovich  density directly by using the Heaviside step-function, which can be easily integrated over, once the roots of its argument are determined.
We remark that the approach of \cite{Pietroni:2018ebj} (applied to a 1D cosmology) also exploits Heaviside step-functions, which are then used as ``force input'' to solve the underlying evolution equation by numerical integration.

Apart from the different force computation of \cite{Taruya:2017ohk}, there are more differences as regards to solving the underlying evolution equations. While
our expansion scheme solves in the first iteration
\be \label{eq:expansionrep}
   \partial_q \left[ \tau^2 \partial_\tau^2 + \frac{3\tau}{2} \partial_\tau - 3/2 \right] \xi_\PSC = - \frac 3 2 F_\ZA \,,
\ee
with $F_\ZA = F(\xofqtextZA)$, \cite{Taruya:2017ohk} solve instead 
\be
  \partial_q \left[ \tau^2 \partial_\tau^2 + \frac{3\tau}{2} \partial_\tau  \right] \tilde \xi_\PSC = - \frac 3 2 \left( F_\ZA - \partial_q \xi_\ZA  \right)
\ee
in our notation, where we denote the resulting PSC displacement of \cite{Taruya:2017ohk} with $\tilde \xi_\PSC$. 
As a consequence, the expansion schemes are fundamentally different, however most likely they will asymptote to the same (convergent) answer at very large orders. 
The truncation errors at low orders, however, could be substantially different and should be assessed in future work. 
Nonetheless, the approach of \cite{Taruya:2017ohk} appears to be suitable to identify all singularities, but only for the symmetric collapse ($c=0$).

We remark that $F_\ZA=0$ in single-stream regions, thus the evolution equation~\eqref{eq:expansionrep} reduces to the standard linear ODE for the ZA displacement with
the temporal operator $\RT = \tau^2 \partial_\tau^2+({3\tau}/2)\partial_\tau\!-3/2$, which has eigenvalues +1 and -3/2,
thus leading to 
the standard growing ($\propto \tau^{+1}$) and decaying-mode ($\propto \tau^{-3/2}$) solutions. 
This identification allows us to solve equation~\eqref{eq:expansionrep} by applying the method of variation of constants, which involves only a single temporal integration (see App.\ \ref{app:xiPSC} for details).
By contrast, in the work of \cite{Taruya:2017ohk}, two temporal integrations are performed, 
for which the authors impose further Taylor expansions.

Finally, we note that the instructions of \cite{Taruya:2017ohk} for higher-order refinements differ from ours. Specifically,  \cite{Taruya:2017ohk} suggest potential improvements that involve higher Taylor approximations (in time and space) of the positions $q_{\rm min/max}$ and $q_{\rm comin/comax}$ (cf. Fig.\,\ref{fig:plot-one}), and/or refining some integral expressions that previously were truncated at the leading order in a temporal expansion. By contrast, our scheme {\it bootstraps} the higher-order displacement to the force computation from the previous order (see equations~\ref{app:expansioschemePP}), which is naturally accompanied by a shift of $q_{\rm min/max}$ and $q_{\rm comin/comax}$ at increasing orders; see Fig.\,\ref{fig:funnel}  and App.\,\ref{app:forcePPSC} for details.

\paragraph*{Post-shell-crossing approaches in the $c\neq0$ case.}
Although the present implementation of the approach of \cite{Taruya:2017ohk} only handles a perfect symmetric collapse, it appears to us that only a few additions are required to {\it generalize it to the asymmetric collapse}.
For this note that in their Green's approach an integration constant  has been set to zero stemming from neglecting the $k=0$ Fourier mode when solving for the Poisson equation (their equation~A5). In the asymmetric case, this integration constant takes generally nonzero values after shell-crossing and thus should be restored.
Next, one needs to determine the force in the presence of asymmetries in the initial conditions, which we parametrize with $c$; for small $c$, the updated force can be determined using perturbation theory (see App.\,\ref{app:forcenoc}). Once the force is updated accordingly, the integration constant $\xi_{\rm c}(\tau)$ can be fixed along the lines as discussed in the main text (see around equation~\ref{heckODE}).

Finally, the approach of~\cite{Pietroni:2018ebj} solves the 1D Vlasov--Poisson equations~\eqref{EOM} together with the Dirac-delta expression for the density (equation~\ref{eq:density}) by direct numerical integration. They also consider the case of adding an asymmetric distortion at the level of the initial conditions (called ``the feature'' in \citealt{Pietroni:2018ebj}). While this general strategy is analogous to our $c\neq0$ case, their initial conditions used rely on an assumed
Gaussian shape.  
Furthermore, to avoid trivial violations of  the center-of-mass condition~\eqref{gauge}, \cite{Pietroni:2018ebj} is not adding spatial counter terms in the initial conditions,
but instead  adjusting a space-independent constant. As a consequence, their (numerical) results are not directly comparable to ours.

\section{Late-time results}\label{app:late-time}

\subsection{Late-time behaviour of the boost}

\setcounter{figure}{3} 

\begin{figure}
 \begin{center}
    \includegraphics[width=0.44\textwidth]{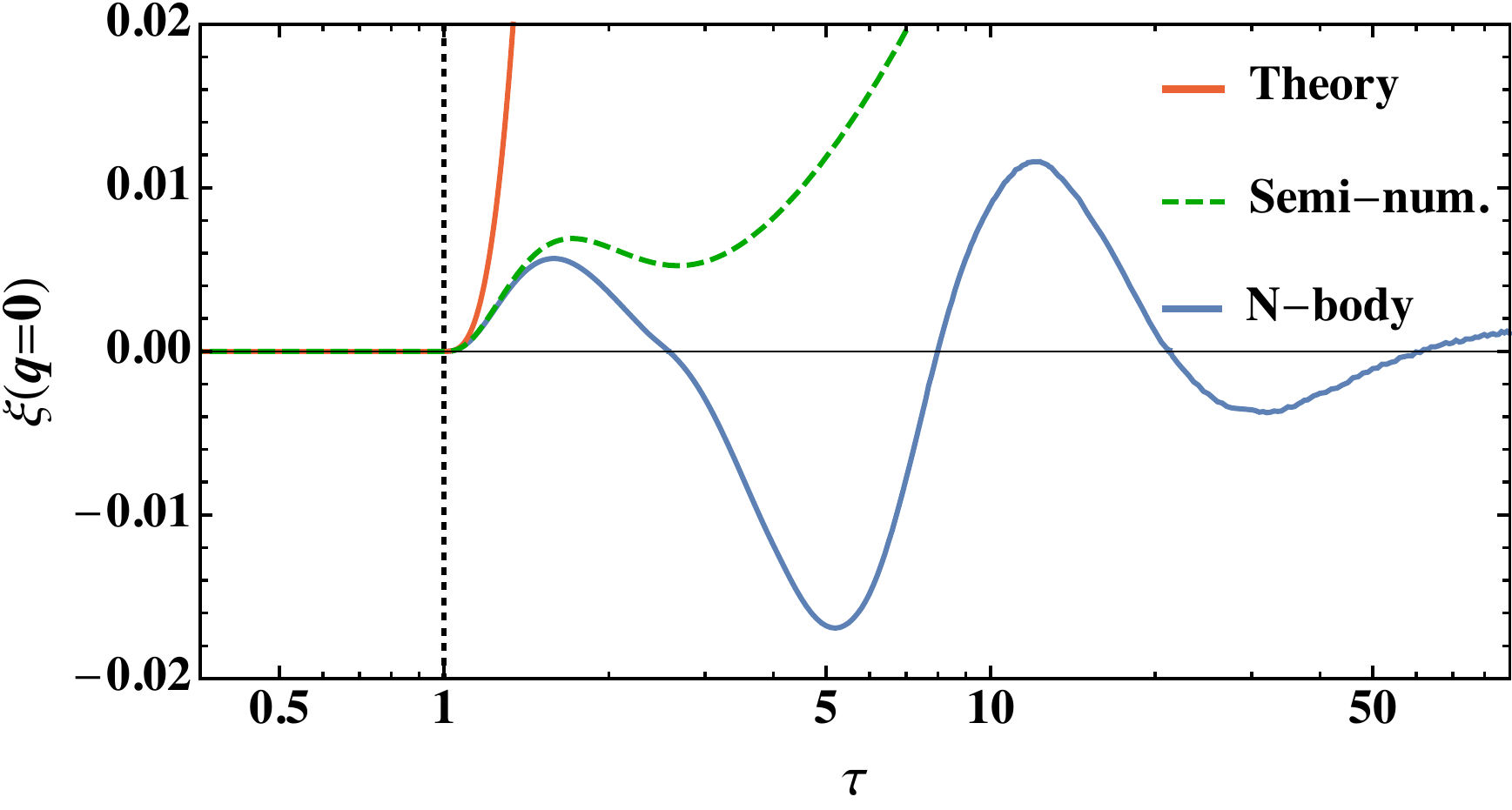}
 \end{center}
\vskip-0.3cm \caption{Evolution of the particle that is initially at position $q=0$, as predicted in leading-order theory (red line), by semi-numerical avenues using suitable approximations (green dashed line; see around equation~\ref{eq:ODEheck}), and from  high-resolution N-body simulations (blue line). We remark that the shown times are well after the first shell-crossing (indicated by a vertical dotted line in black; the second shell-crossing occurs at $\tau \simeq 2.4$). 
The asymmetry parameter  in the initial conditions is $c=0.1$, and for the simulations we used $10^5$ N-body particles and $10^5$ time steps.}
 \label{fig:latetime1}
\end{figure}

As we have seen the boost $\xi_{\rm c}(\tau)$ appears at shell-crossing when the collapse is not perfectly symmetric (parametrized with the asymmetry parameter~$c$).
One consequence of this boost is that the particle that is initially at the position $x=q=0$ begins moving after the first shell-crossing.
From previous considerations (see especially Appendix~\ref{app:forcenoc}) the evolution equation of the particle that is initially at $q=0$ is, exactly and at all times,
\be \label{eq:ODEheck}
  \RT \xi_{\rm c} = \frac 3 2 \left[ \uppi  - \int  \Theta \left( -\xofqprime \right)   \dd q' \right] \,,
\ee
where the factor of $\uppi$ is a boundary term that can be read off from equation~\eqref{eq:SZA1app}.  
Figure~\ref{fig:latetime1} shows the late-time behaviour of that particle as predicted from the leading-order theory (red line; see equation~\ref{heck}), 
from high-resolution numerical simulations (blue line), as well as by numerically integrating equation~\eqref{eq:ODEheck} with the replacement $x(q',\tau) \to x_\ZA(q',\tau)$ on its RHS (green dashed line). We remark that for the two latter avenues we use $v^{\rm (ini)}$ as given in equation~\eqref{IC}, while the leading-order theory exploits normal-form considerations (using equation~\ref{eq:vininormal} for $v^{\rm (ini)}$).

From Fig.\,\ref{fig:latetime1} it is clear, that the leading-order theory can not encapsulate the non-trivial late-time evolution as observed in the simulation. This is not surprising, as the underlying (analytical) computation is expected to be only meaningful shortly after the first shell-crossing; higher-order refinements as well as theoretical extensions beyond the {\it second} shell-crossing are likely to change the conclusions significantly. We keep such avenues for future investigations.

\begin{figure}
 \begin{center}
    \includegraphics[width=0.44\textwidth]{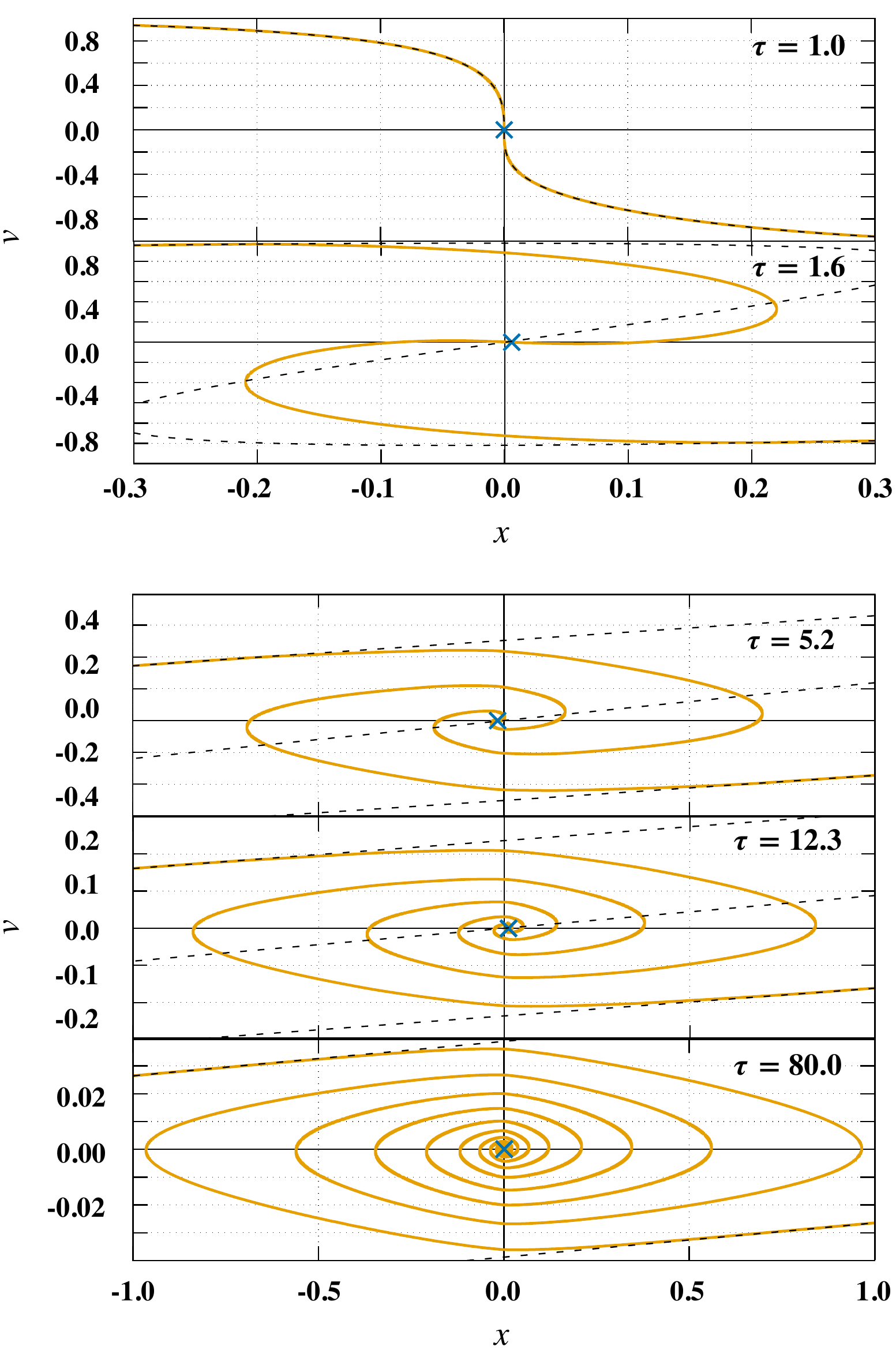}
 \end{center}
\vskip-0.3cm \caption{Numerical solution (solid orange lines) of the phase-space for various times in the case of $c=0.1$, using $10^5$ N-body particles and $10^5$ time steps. The blue cross denotes the phase-space position of the particle that is at $x=0$ until shell-crossing ($\tau=1$), which then begins moving due to the nontrivial forcing asymmetry (see also Fig.\,\ref{fig:latetime1}). The black dashed lines denote results in the ZA. Note the different axes scalings employed.}
 \label{fig:latetime2}
\end{figure}

In Fig.\,\ref{fig:latetime2} we show the corresponding phase-space for various times, where the current position of the fluid particle with label $q=0$  is marked with a blue cross (i.e., the corresponding spatial positions coincide with those from Fig.\,\ref{fig:latetime1} at the respective times).
Similarly as in the previous figure, we observe a damped oscillating behaviour of that particle for sufficiently late times; only at very late times $\tau >1000$, we find that
the absolute deviation from zero of this particle is less than $10^{-4}$ (not shown). Naturally, in the limit $\tau \to \infty$ one would expect that $ \xi(q=0) \to 0$, which is however difficult to achieve in numerical implementations. We keep the study of such attractors for future work.

\begin{figure*}
 \begin{center}
    \includegraphics[width=0.95\textwidth]{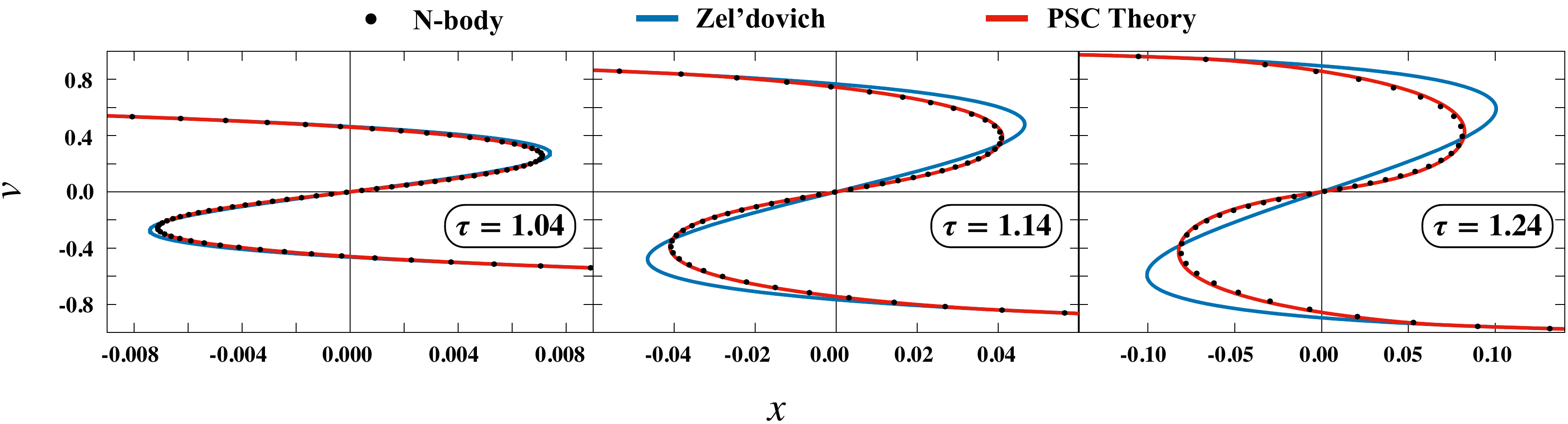}
 \end{center}
 \caption{Late-time results in ZA and PSC theory shown respectively in red and blue solid lines, compared against results from our numerical simulations (using 8192 particles/time steps) indicated with black dots.  For reasons of better visibility, only a reduced amount of data points from the simulation  is shown (left to right panel: every 30th, 50th and 80th data point). Here we have set $c=0$ and, as before, computations are performed on the torus with a period~$x \in [-\uppi,\uppi]$ (i.e., only the interesting parts of the whole phase-space are shown).}
 \label{fig:latetime3}
\end{figure*}

\subsection{Comparison between ZA and PSC theory}

For completeness, we  added the ZA prediction in Fig.\,\ref{fig:latetime2}, shown as black dashed lines. Clearly, the ZA is unable to predict the two essential features in the multi-stream regime, namely (1) the development of confined multi-stream regions (which is well known, see e.g.\ \citealt{Taruya:2017ohk}), and (2) the non-trivial (right/left) shift of the overall multi-streaming region due to the force imbalance, inherent to asymmetric collapse scenarios ($c \neq 0$) which, as we elucidate in the main text, has not been investigated before in the literature.

Finally,
to eludicate the aforementioned feature (1) further, we show in Fig.\,\ref{fig:latetime3} a comparison between ZA (blue solid line), PSC theory (red solid line) and simulations (black dots) at late times for the case of $c=0$ (note the different $x$-scalings employed in the panels). While the results from the PSC theory agree reasonably well with those from the numerical simulations until $\tau \lesssim 1.24$, the ZA exemplifies the common overshooting problem, i.e., it lacks the ability for predicting secondary gravitational infall. We remark that this inability of the ZA is well documented in the literature; for recent in-depth comparisons at (even) later times between ZA and post-shell-crossing theories/simulations, see \cite{Taruya:2017ohk,Pietroni:2018ebj}.

\vfill

\bsp	
\label{lastpage}
\end{document}